\documentclass[aps,prd,twocolumn,floatfix,superscriptaddress,longbibliography,10pt]{revtex4-2}

\usepackage[latin9]{inputenc}
\usepackage{amsmath}
\usepackage{amssymb}
\usepackage{graphicx}
\usepackage{esint}
\usepackage{epstopdf}
\usepackage{braket}
\usepackage{bm}
\usepackage{xcolor}
\usepackage[colorlinks=true,citecolor=blue,linkcolor=blue,urlcolor=blue]{hyperref}
\usepackage{bibunits}
\usepackage{mathtools}
\usepackage{stackengine}
\usepackage{dsfont}
\usepackage{soul}
\usepackage{multirow}
\usepackage{placeins}
\usepackage{xspace}
\usepackage{nicefrac}
\usepackage{tikz}
\usepackage{verbatim}
\usetikzlibrary{arrows,shapes}
\usetikzlibrary{automata,positioning}
\usetikzlibrary{positioning}
\usetikzlibrary{angles,quotes}

\newcommand{\trans}{\ensuremath{^{\textrm{T}}}\xspace}

\begin{document}

\title{Nonlocal advantage of quantum coherence in tau-lepton pairs from electron--positron collisions}

\author{Yoav Afik}
\email{yoavafik@gmail.com}
\affiliation{Enrico Fermi Institute, University of Chicago, Chicago, Illinois 60637, USA}

\author{Juan Ram\'on Mu\~noz de Nova}
\email{jr.denova@csic.es}
\affiliation{Instituto de Estructura de la Materia, IEM-CSIC, Serrano, 123 E-28006 Madrid, Spain}

\author{Sarthak Sharma}
\email{b23ph1020@iitj.ac.in}
\affiliation{Indian Institute of Technology Jodhpur, Jodhpur 342037, India}

\author{Surya Sundar Raman}
\email{surya.sundar.raman@cern.ch}
\affiliation{Department of Physics, University of British Columbia, Vancouver BC, Canada}
\affiliation{TRIUMF, 4004 Wesbrook Mall, Vancouver, BC V6T 2A3, Canada}

\begin{abstract}
Quantum correlations have recently attracted significant attention in collider physics, with Nonlocal Advantage of Quantum Coherence (NAQC) representing the strongest form of quantum correlation studied so far.
Owing to its strength, NAQC is considerably more challenging to observe than Bell nonlocality since it is only present in narrower regions of phase space. 
Unlike other systems, tau-lepton pair ($\tau^+\tau^-$) production in electron--positron ($e^+e^-$) collisions provides an ideal testing ground, as it exhibits strong quantum correlations and, in particular, NAQC over a large region of phase space.
In this paper, we investigate NAQC in $e^+e^- \to \tau^+\tau^-$ production at center-of-mass energies of 10.58 and 91.19~GeV, corresponding to the ongoing Belle~II and future FCC-$ee$ experiments, respectively.
Remarkably, for our experimental proposal, we develop a dedicated NAQC witness, which allows to certify NAQC presence by measuring one single magnitude. 
We estimate the experimental sensitivities needed to establish the presence of NAQC, finding that a large fraction of the number of events (specifically, 0.18 for Belle~II and 0.31 for FCC-$ee$) contribute to the NAQC signal. 
A 5$\sigma$ observation of NAQC requires a precision of approximately 1\% at Belle~II, while a precision of a few percent is sufficient at FCC-$ee$. 
The resulting observation would constitute the strongest form of quantum correlation measured in a collider to date.
In addition, we design an experimental scheme to implement the steering game leading to the NAQC definition, representing the first implementation of a steering game in a high-energy collider. 
Our results are easily extendable to other electron--positron colliders with different center-of-mass energies. 
\end{abstract}

\maketitle

\tableofcontents

\section{Introduction}\label{sec:Intro}

Quantum correlations in fundamental processes at particle colliders have attracted significant attention in recent years. 
These correlations, including quantum discord, entanglement, steering, and Bell nonlocality, have been investigated in a wide variety of systems, including top--antitop quark pairs~\cite{Afik:2020onf,Fabbrichesi:2021npl,Afik:2022kwm,Severi:2021cnj,Aguilar-Saavedra:2022uye,Afik:2022dgh,Dong:2023xiw,Cheng2023,Aguilar-Saavedra:2024hwd,Aguilar-Saavedra:2024fig,Han:2023fci,White:2024nuc,Cheng:2024btk,Han:2024ugl,Fabbrichesi:2025rsg,Fabbrichesi:2025psr,Martinez-Moreno:2025jgx}, heavy gauge bosons~\cite{Barr:2021zcp,Barr:2022wyq,Ashby-Pickering:2022umy,Aguilar-Saavedra:2022wam,Aguilar-Saavedra:2022mpg,Bernal:2023ruk,Morales:2023gow,Fabbri:2023ncz,Fabbrichesi:2023jep,Aguilar-Saavedra:2024whi,Bernal:2024xhm,Grossi:2024jae,Goncalves:2025mvl}, bottom--antibottom quark pairs~\cite{Kats:2023zxb,Afik:2025grr}, neutrinos~\cite{Blasone:2007vw,Formaggio:2016cuh,Ming2020,Blasone:2021cau}, and tau-lepton pairs ($\tau^+\tau^-$)~\cite{Fabbrichesi:2022ovb,Altakach:2022ywa,Ehataht:2023zzt,Fabbrichesi:2024wcd,Han:2025ewp,Zhang:2025mmm,Yang:2026uwu,Fang:2026ddi}. More comprehensive reviews of these developments can be found in Refs.~\cite{Barr:2024djo,Afik:2025ejh}. Several of these theoretical proposals have already culminated in experimental measurements of quantum entanglement in top--antitop quark pairs~\cite{ATLAS:2023fsd,CMS:2024pts,CMS:2024zkc,CMS2026} and in $Z$ bosons~\cite{ATLAS:2026hye}, performed by the ATLAS and CMS collaborations at the LHC. 
At the same time, these developments have raised fundamental discussions regarding the interpretation of quantum information measurements in high-energy collider experiments, summarized recently by the ATLAS collaboration in Ref.~\cite{ATL-PHYS-PUB-2026-006}.

Among these quantum correlations, a well-defined hierarchy exists. In particular, for bipartite systems, such as pairs of top--antitop quarks, heavy gauge bosons, or tau leptons, this hierarchy can be summarized as follows~\cite{Wiseman2007,Qureshi2018,Baker2020,Uola2020}:
\begin{equation*}
\begin{aligned}
\textrm{Bell Nonlocality} \subset \textrm{Steering} \subset 
 \textrm{Entanglement} \subset \textrm{Discord}.
\end{aligned}
\end{equation*}
For instance, every Bell nonlocal state is necessarily steerable, entangled, and exhibits nonzero quantum discord. 
The converse, however, does not generally hold; for example, an entangled state does not have to violate a Bell inequality.
The hierarachy of such correlations in a high-energy two-qubit system has been studied in Ref.~\cite{Afik:2022dgh}, and demonstrated using experimental data in Ref.~\cite{Afik:2026pxv}, for the case of a top-antitop system.

Another nonlocal manifestation of quantum correlations is the Nonlocal Advantage of Quantum Coherence (NAQC)~\cite{Mondal2017}.
A bipartite two-qubit system exhibits NAQC if one party, by performing local measurements and communicating the outcomes, can remotely prepare conditional states for the other party whose average quantum coherence exceeds the maximum coherence achievable by any single local quantum state. 
As a result, NAQC is a particularly strong manifestation of quantum correlations, yielding an operational advantage in the generation of coherence.
Indeed, NAQC is a sufficient condition for entanglement and steerability~\cite{Mondal2017} and, for states with diagonal correlation matrix, also for Bell nonlocality~\cite{Hu2018}. 
A general proof that NAQC implies Bell nonlocality is still missing, although there is strong numerical evidence pointing in this direction~\cite{Hu2018}.
Furthermore, NAQC is asymmetric, in the same fashion as quantum discord and steering~\cite{Ollivier2001,Wiseman2007}. 
The concept of NAQC can be generalized to pairs of qudits using mutually unbiased bases when their dimension $d$ is the power of a prime number~\cite{Hu2018Dimension}. 
Experimentally, NAQC has been observed in a photonic setup~\cite{Ding2019}.

In the high-energy context, NAQC has been studied in neutrinos~\cite{Blasone:2021cau,Bittencourt:2022tcl,Yadav:2022grk,Blasone:2022ete} and top-antitop quark pairs at the LHC~\cite{Afik:2022dgh,Rai:2025qke}. 
In the latter case, NAQC is only present in highly confined regions of phase space, whose measurement is unrealistic in any foreseeable future.
In particular, Ref.~\cite{Rai:2025qke} showed that no NAQC signal is present in the current experimental measurements of top--antitop quark pairs. 
More generally, the study of coherence in high-energy colliders is currently attracting a lot of attention~\cite{Larkoski2022,Gu:2025ijz,Haddadi:2026nyg,Agrawal:2026zwa,Coci:2026puh,Aoude:2026eeg,Cheng:2026zfb}.

In order to experimentally probe a certain form of quantum correlations in a high-energy collider, it is advantageous to consider processes in which such correlations persist over a large portion of phase space, rather than being confined to small kinematic regions. 
A particularly suitable environment is provided by $\tau^+\tau^-$ production in $e^+e^-$ collisions, where the center-of-mass (COM) energy of the process is precisely known and the produced tau leptons exhibit strong quantum correlations.

The tau lepton is the heaviest lepton in the Standard Model, with a mass of $m_\tau \approx 1.78$~GeV and a lifetime of $\tau_\tau \approx 2.9 \times 10^{-13}$~s~\cite{ParticleDataGroup:2026mpi}. 
Unlike quarks, tau leptons do not participate in strong interactions and therefore do not hadronize. 
Consequently, their spin information is preserved until decay and is encoded in the angular distributions of their decay products, enabling an accurate reconstruction of their spin quantum state. 
Large samples of $\tau^+\tau^-$ pairs are already available at experiments such as Belle~\cite{Belle:2012iwr}, Belle~II~\cite{Belle-II:2018jsg} and BESIII~\cite{BESIII:2009fln}, and will be produced in even greater numbers at proposed future facilities including FCC-$ee$~\cite{FCC:2018evy}, LEP~III~\cite{Blondel:2012ey}, CEPC~\cite{CEPCStudyGroup:2023quu}, and proposed linear colliders~\cite{Behnke:2013xla,CLIC:2016zwp,Nanni:2023yne}.

In this work, we analyze the presence of NAQC in $\tau^+\tau^-$ from $e^+e^-$ collisions.
We focus on two benchmark cases where $\tau^+\tau^-$ pairs are generated with large statistics: electron--positron collisions at a COM energy of 10.58~GeV, such as in Belle~\cite{Belle:2012iwr} and Belle~II~\cite{Belle-II:2018jsg}, and 91.19~GeV, as expected in FCC-$ee$~\cite{FCC:2018evy}. Nevertheless, our analysis can be easily extended to $e^+e^-$ colliders operating at other COM energies.
Remarkably, we construct an NAQC witness (based on existing entanglement witnesses~\cite{Afik:2020onf,Afik:2022kwm,Aguilar-Saavedra:2022uye}) which allows to signal the presence of NAQC from the measurement of a single magnitude.
Our experimental assessment finds that a large fraction of the number of events (in particular, 0.18 for Belle~II and 0.31 for FCC-$ee$) contribute to the NAQC signal, potentially allowing the observation of NAQC, which would represent the strongest form of quantum correlation measured in a collider to date. 
Specifically, we find that a 5$\sigma$ observation of NAQC requires a precision of approximately 1\% at Belle~II, and of a few percent at FCC-$ee$. 
Finally, we design an experimental scheme to implement the steering game leading to the NAQC definition, constituting the first implementation of a steering game~\cite{Cavalcanti2013,Banik2013,Sun2014,Kocsis2015,Gheorghiu2017} in a high-energy collider and paving the way for more general steering games in the future.

The article is arranged as follows. 
Section~\ref{sec:Theory} discusses the theoretical framework upon which our results are built, including the concept of NAQC and the electroweak theory describing the spin quantum state in $e^+e^- \to\tau^+\tau^-$ production. 
Our study of NAQC is presented in Section~\ref{sec:Results}, and our experimental proposal is described in Section~\ref{sec:Experimental}.
Conclusions and outlook are provided in Section~\ref{sec:Conclu}. Appendices contain technical details behind the results of the work.

\section{Theoretical framework}\label{sec:Theory}

\subsection{Coherence}\label{subsec:coherence}

We begin by reviewing the quantitative characterization of the coherence of a quantum state. 
Intuitively, coherence denotes the fact that quantum states can be written as superpositions of states which possess well-defined relative phases. 
For a pure state $\ket{\psi}$ in a Hilbert space $\mathcal{H}$ of dimension $N$, we can say that $\ket{\psi}$ is coherent in a certain basis $\{\ket{i}\}^N_{i=1}$ when the expansion
\begin{equation}
\ket{\psi}=\sum^N_{i=1} c_i\ket{i}
\end{equation}
contains more than one non-zero coefficient. 
Within the more general density-matrix formalism, a quantum state is coherent when the expansion of its density matrix
\begin{equation}
\rho=\sum^N_{i,j=1}\rho_{ij}\ket{i}\bra{j}
\end{equation}
contains off-diagonal terms. 
Equivalently, a quantum state is \textit{incoherent} when its density matrix is diagonal,
\begin{equation}
\label{eq:Diagonal}
\rho=\rho_D=\sum^N_{i=1} \rho_{ii}\ket{i}\bra{i}.
\end{equation}
The presence of coherence is revealed when measuring an appropriate observable $O$ sensitive to those off-diagonal terms, since its expectation value can be written as
\begin{equation}
\braket{O}=\textrm{Tr}[O\rho]=\sum^N_{i=1}\rho_{ii}\braket{i|O|i}+\sum^N_{i\neq j}\rho_{ij}\braket{j|O|i}.
\end{equation}
Thus, coherence results in interference terms $\sim \braket{j|O|i}$, which are sensitive to the individual phase of each state. 
As a result, coherence is naturally a basis-dependent property, where the specific basis is selected by the system of interest.

A more quantitative and precise characterization of the concept of coherence was developed in Ref.~\cite{Baumgratz2014}. 
Among other properties, any good measure of coherence $\mathfrak{C}$ should satisfy i) $\mathfrak{C}(\rho)=0$ \textit{iff} $\rho$ is incoherent; and ii) convexity, i.e., it is non-increasing under mixing. 
Specifically, convexity means that for a convex sum of quantum states
\begin{equation}
\label{eq:convexsum}
\rho=\sum_{n}p_n\rho_n,~p_n\geq 0,~\sum_np_n=1,
\end{equation}
the coherence measure satisfies
\begin{equation}
\label{eq:convexity}
\mathfrak{C}(\rho)=\mathfrak{C}\left(\sum_{n}p_n\rho_n\right)\leq \sum_{n}p_n \mathfrak{C}\left(\rho_n\right).
\end{equation}
This property implies that maximally coherent states are pure states (notice that any mixed state can be always written as a convex sum of the pure states given by the eigenstates of its density matrix). 
Specifically, it can be seen~\cite{Baumgratz2014} that a pure state of the form
\begin{equation}
\label{eq:MaximallyCoherent}
\ket{\psi}=\sum^N_{i=1}c_i\ket{i},~|c_i|=\frac{1}{\sqrt{N}},
\end{equation}
is always maximally coherent.

A simple and good measure of coherence is given by the $l_1$-norm of coherence (LNC)~\cite{Baumgratz2014}:
\begin{equation}
\label{eq:l1norm}
\mathfrak{C}^{l_1}(\rho)\equiv \sum^N_{i\neq j} |\rho_{ij}|.
\end{equation}
Another possible measure is given by the relative entropy of coherence (REC)~\cite{Baumgratz2014}:
\begin{equation}
\label{eq:re}
\mathfrak{C}^{\rm{re}}(\rho)\equiv S(\rho_D)-S(\rho),
\end{equation}
where $\rho_D$ is the diagonal part of $\rho$, Eq.~(\ref{eq:Diagonal}), and $S(\rho)=-\textrm{Tr}(\rho\log_2\rho)$ is the von Neumann entropy.
Again, we stress that these are basis-dependent magnitudes, as their values depend on the specific orthonormal basis of the Hilbert space in which the matrix elements $\rho_{ij}$ of $\rho$ are evaluated. 
It is easy to show that the maximally coherent state (\ref{eq:MaximallyCoherent}) maximizes both LNC and REC measures.

For illustrative purposes, we consider the case of a single qubit (i.e., a two-level quantum system), where the density matrix can be simply expanded as
\begin{equation}
\rho=\frac{I_2+\sum_{i}B_{i}\sigma^i}{2}=\frac{1+\mathbf{B}\cdot\boldsymbol{\sigma}}{2},
\end{equation}
with $I_n$ the $n\times n$ identity matrix and $\sigma^i$ the Pauli matrices. Without loss of generality, we can always think of a qubit as a spin-1/2 particle. 
In that case, the vector $\mathbf{B}$ (usually referred to as the Bloch vector) represents the spin polarization of the particle,
\begin{equation}
\mathbf{B}=\braket{\boldsymbol{\sigma}}=\textrm{Tr}[\rho\boldsymbol{\sigma}].
\end{equation}
The eigenstates and eigenvalues of $\rho$ are
\begin{equation}
\rho\ket{\pm \mathbf{\hat{n}}_B}= p_{\pm}\ket{\pm \mathbf{\hat{n}}_B},~p_{\pm}=\frac{1\pm|\mathbf{B}|}{2},
\end{equation}
$\ket{\pm \mathbf{\hat{n}}}$ being 
the eigenstates of the spin projection along a certain direction $\mathbf{\hat{n}}$, $(\mathbf{\hat{n}}\cdot\boldsymbol{\sigma})\ket{\pm \mathbf{\hat{n}}}=\pm \ket{\pm \mathbf{\hat{n}}}$, and $\mathbf{\hat{n}}_B$ being the direction of $\mathbf{B}$, $\mathbf{B}=|\mathbf{B}|\mathbf{\hat{n}}_B$.

Physical states (i.e., those described by non-negative density matrices) correspond to Bloch vectors $\mathbf{B}$ contained inside the unit sphere, $|\mathbf{B}|\leq 1$, which is then denoted as the Bloch sphere. 

In the usual basis $\ket{\pm}$ of the eigenstates of $\sigma_z$, $\sigma_z\ket{\pm}=\pm \ket{\pm}$, $\rho$ can be explicitly written as
\begin{equation}
\label{eq:GeneralQubitExplicit}
\rho=\frac{1}{2}\left[\begin{array}{cc}
1+B_3 & B_{1}-iB_2 \\
B_{1}+iB_2 & 1-B_3 \\
\end{array}\right].
\end{equation}
In this basis, the LNC and REC simply read
\begin{align}
\mathfrak{C}^{l_1}(\rho)&=\sqrt{B^2_1+B^2_2},\\
\nonumber \mathfrak{C}^{\rm{re}}(\rho)&=h\left(\frac{1+B_3}{2}\right)-h\left(\frac{1+|\mathbf{B}|}{2}\right),
\end{align}
where $h(x)=-x\log_2 x-(1-x)\log_2(1-x)$ is the binary entropy.

More generally, any orthonormal qubit basis is composed of two spin eigenstates of the form $\ket{\pm \mathbf{\hat{n}}}$. 
Thus, given a certain qubit density matrix $\rho$, any coherence measure $\mathfrak{C}(\rho)$ is fixed just by the spin direction $\mathbf{\hat{n}}$ determining the underlying basis, which is then denoted as $\mathfrak{C}_{\mathbf{\hat{n}}}(\rho)$. Specifically, we have
\begin{align}
\label{eq:DirectionCoherence}
\mathfrak{C}_{\mathbf{\hat{n}}}^{l_1}(\rho)&=|\mathbf{\hat{n}}\times\mathbf{B}|,\\
\nonumber \mathfrak{C}_{\mathbf{\hat{n}}}^{\rm{re}}(\rho)&=h\left(\frac{1+\mathbf{\hat{n}}\cdot\mathbf{B}}{2}\right)-h\left(\frac{1+|\mathbf{B}|}{2}\right).
\end{align}
These measures are bounded as
\begin{align}
\label{eq:CoherenceBound}
0&\leq \mathfrak{C}^{l_1}_{\mathbf{\hat{n}}}(\rho)=\sqrt{|\mathbf{B}|^2-(\mathbf{\hat{n}}\cdot\mathbf{B})^2}\leq |\mathbf{B}|\leq 1,\\
\nonumber 0&\leq \mathfrak{C}_{\mathbf{\hat{n}}}^{\rm{re}}(\rho)\leq h\left(\frac{1+\mathbf{\hat{n}}\cdot\mathbf{B}}{2}\right)\leq h\left(\frac{1}{2}\right)=1.
\end{align}
The maximum coherence in both cases is then achieved for a pure state, $|\mathbf{B}|=1$, whose spin polarization is perpendicular to $\mathbf{\hat{n}}$, $\mathbf{\hat{n}}\cdot\mathbf{B}=0$, in agreement with Eq.~(\ref{eq:MaximallyCoherent}). 
In the case where $\mathbf{\hat{n}}=\hat{z}$, this means that the maximally coherent state is a pure state whose spin polarization is contained in the $xy$-plane.

Given a certain orthonormal basis $\{\mathbf{\hat{n}}_i\}^3_{i=1}$ of the three-dimensional space, we denote the coherence in the Hilbert-space basis $\ket{\pm \mathbf{\hat{n}}_i}$ as $\mathfrak{C}_i(\rho)\equiv \mathfrak{C}_{\mathbf{\hat{n}}_i}(\rho)$. An average coherence over all directions can be then defined as
\begin{equation}
\bar{\mathfrak{C}}(\rho)\equiv\sum^3_{i=1}\mathfrak{C}_{i}(\rho).
\end{equation}
The average LNC is bounded as
\begin{equation}
\label{eq:LNCuality}
\bar{\mathfrak{C}}^{l_1}(\rho)=\sum^3_{i=1}\mathfrak{C}^{l_1}_{i}(\rho)=\boldsymbol{\mathfrak{C}}\cdot\mathbf{1}\leq |\boldsymbol{\mathfrak{C}}|\cdot|\mathbf{1}|=\sqrt{3\cdot }|\boldsymbol{\mathfrak{C}}|\leq \sqrt{6},
\end{equation}
where $\boldsymbol{\mathfrak{C}}=[\mathfrak{C}^{l_1}_1,\mathfrak{C}^{l_1}_2,\mathfrak{C}^{l_1}_3]$, $\mathbf{1}=[1,1,1]$, we have used the Cauchy-Schwarz inequality, and
\begin{align}
\nonumber |\boldsymbol{\mathfrak{C}}|^2&=\sum^3_{i=1}\left[\mathfrak{C}^{l_1}_{i}(\rho)\right]^2=\sum^3_{i=1}|\mathbf{\hat{n}}_i\times\mathbf{B}|^2\\
&=\sum^3_{i=1}\left[|\mathbf{B}|^2-(\mathbf{\hat{n}}_i\cdot \mathbf{B})^2\right]=2|\mathbf{B}|^2\leq 2.
\end{align}
The inequality (\ref{eq:LNCuality}) is then saturated for a pure state satisfying $\boldsymbol{\mathfrak{C}}\propto \mathbf{1}$, which implies an isotropic polarization vector 
\begin{equation}
\label{eq:MaximumQubit}
\mathbf{B}=\pm \dfrac{\mathbf{1}}{\sqrt{3}}.
\end{equation}

For the average REC, we have
\begin{align}
\nonumber \bar{\mathfrak{C}}^{\rm{re}}(\rho)&=\sum^3_{i=1}\left[h\left(\frac{1+\mathbf{\hat{n}}_i\cdot \mathbf{B}}{2}\right)-h\left(\frac{1+|\mathbf{B}|}{2}\right)\right]\\
&=\sum^3_{i=1}h\left(\frac{1+\mathbf{\hat{n}}_i\cdot \mathbf{B}}{2}\right)-3h\left(\frac{1+|\mathbf{B}|}{2}\right).
\end{align}
Due to the symmetry of this expression, its maximum is again obtained for the isotropic pure state of Eq.~(\ref{eq:MaximumQubit}). Hence, the average LNC and REC for a single qubit are bounded as
\begin{align}
\label{eq:MaxQubit}
\bar{\mathfrak{C}}^{l_1}(\rho)&\leq \bar{\mathfrak{C}}^{l_1}_{\rm{max}}=\sqrt{6}\approx 2.45,\\
\nonumber \bar{\mathfrak{C}}^{\rm{re}}(\rho)&\leq \bar{\mathfrak{C}}^{\rm{re}}_{\rm{max}}=3h\left(\frac{1+\frac{1}{\sqrt{3}}}{2}\right)\approx 2.23.
\end{align}

\subsection{NAQC}\label{subsec:NAQC}

The situation gets more interesting when we consider a bipartite system, $\mathcal{H}=\mathcal{H}_A\otimes \mathcal{H}_B$, where the subsystems $A,B$ are referred to as Alice and Bob, respectively. Once more, for illustrative purposes, we consider a two-qubit system, whose density matrix is now expanded as
\begin{equation}
\label{eq:densitymatrix2qubit}
\rho= \frac{I_4 + \sum_i (B^+_i \sigma_i \otimes I_2 \
+B^-_i I_2 \otimes \sigma_i) + \sum_{ij} C_{ij} \sigma_i \otimes \sigma_j}{4}.
\end{equation}
The coefficients $B^{\pm}_i$ encode the spin polarizations $\mathbf{B}^\pm$ of Alice and Bob, respectively, while $C_{ij}$ represent the elements of the spin-correlation matrix $\mathbf{C}$,
\begin{equation}
\label{eq:SpinCoefficients}
B^{+}_{i}=\braket{\sigma^i\otimes I_2},~B^{-}_{i}=\braket{I_2\otimes\sigma^i},~C_{ij}=\braket{\sigma^{i}\otimes\sigma^{j}}.
\end{equation}
As well known, bipartite systems can give rise to quantum correlations, such as entanglement, which go beyond those achievable in classical statistics. A particularly interesting concept is that of steering~\cite{Wiseman2007}, in which Alice/Bob can remotely manipulate the state of Bob/Alice by performing local measurements on their side. Specifically, if Bob performs a projective measurement of the spin polarization along the direction $\mathbf{\hat{n}}$, he will find his qubit in the state $\ket{\mathbf{\hat{n}}}$ with probability
\begin{equation}
\label{eq:ConditionalProbability}
p_\mathbf{\hat{n}}=\textrm{Tr}\left[(I_2\otimes \Pi_\mathbf{\hat{n}})\rho(I_2\otimes \Pi_\mathbf{\hat{n}})\right]=\frac{1+\mathbf{\hat{n}}\cdot\mathbf{B}^-}{2},
\end{equation} $\Pi_\mathbf{\hat{n}}=\ket{\mathbf{\hat{n}}}\bra{\mathbf{\hat{n}}}$ being the projector onto the state $\ket{\mathbf{\hat{n}}}$. The postmeasurement one-qubit state of Alice is then
\begin{equation}
\label{eq:Postate}
\rho_\mathbf{\hat{n}}= \frac{\textrm{Tr}_B\left[(I_2\otimes \Pi_\mathbf{\hat{n}})\rho(I_2\otimes \Pi_\mathbf{\hat{n}})\right]}{p_\mathbf{\hat{n}}}=\frac{1+\mathbf{B}^+_{\mathbf{\hat{n}}}\cdot\boldsymbol{\sigma}}{2},
\end{equation}
with
\begin{equation}
\label{eq:ConditionalBloch}
\mathbf{B}^+_{\mathbf{\hat{n}}}=\frac{\mathbf{B}^+ +\mathbf{C}\cdot\mathbf{\hat{n}}}{1+\mathbf{\hat{n}}\cdot\mathbf{B}^-}.
\end{equation}
The set of all possible conditional Bloch vectors $\mathbf{B}^+_{\mathbf{\hat{n}}}$ (as a function of $\mathbf{\hat{n}}$) for a given quantum state defines the surface of the so-called steering ellipsoid~\cite{Jevtic2014}, which provides a useful geometrical tool to represent two-qubit states, extending the one-qubit concept of the Bloch sphere.

The steerability of Alice qubit has deeper implications, as it may allow Bob to remotely manipulate Alice coherence beyond the locally achievable limits; this is the idea behind the concept of NAQC. 
The presence of NAQC can be revealed by the following steering game~\cite{Mondal2017}: Bob performs a projective spin measurement along the axis $j$ of a certain orthonormal basis, obtaining the outcome $a=\pm 1$ with probability
\begin{equation}
\label{eq:ConditionalProbabilityNAQC}
p^{a}_j\equiv p_{a\mathbf{\hat{n}}_j}=\frac{1+a\mathbf{\hat{n}}_j\cdot\mathbf{B}^-}{2}.
\end{equation}
The conditional quantum state of Alice after Bob measurement is then
\begin{align}
\label{eq:NAQCBloch}
\nonumber \nonumber \rho^{a}_j&\equiv\rho_{a\mathbf{\hat{n}}_j}=\frac{1+\mathbf{B}^+_{a,j}\cdot\boldsymbol{\sigma}}{2},\\ \mathbf{B}^+_{a,j}&\equiv \mathbf{B}^+_{a\mathbf{\hat{n}}_j}=\frac{\mathbf{B}^+ +a\mathbf{C}\cdot\mathbf{\hat{n}}_j}{1+a\mathbf{\hat{n}}_j\cdot\mathbf{B}^-}.
\end{align}
Now, Alice can measure the coherence, randomly, in the two remaining orthogonal directions $i\neq j$. By averaging the process over all possible orthogonal directions $j$, Bob can claim to have achieved a nonlocal advantage on the quantum coherence of Alice if 
\begin{equation}
\label{eq:NAQCDef}
\mathfrak{C}_{\rm{NA}}(\rho)\equiv\sum^3_{j=1}
\sum^3_{\substack{i=1 \\ i\neq j}}\frac{1}{2}\sum_{a=\pm 1}p^{a}_{j}\mathfrak{C}_{i}(\rho^{a}_{j})
\end{equation}
surpasses the maximum bound of the one-qubit average coherence measure $\bar{\mathfrak{C}}(\rho)$, $\mathfrak{C}_{\rm{NA}}(\rho)>\bar{\mathfrak{C}}_{\rm{max}}$. Specifically, for the average LNC and REC, NAQC is achieved when [see Eq.~(\ref{eq:MaxQubit})]:
\begin{align}
\label{eq:NAQCBound}
\mathfrak{C}_{\rm{NA}}^{l_1}(\rho)&> \bar{\mathfrak{C}}^{l_1}_{\rm{max}}=\sqrt{6},\\
\nonumber \mathfrak{C}_{\rm{NA}}^{\rm{re}}(\rho)&> \bar{\mathfrak{C}}^{\rm{re}}_{\rm{max}}=3h\left(\frac{1+\frac{1}{\sqrt{3}}}{2}\right).
\end{align}
On the other hand, from Eq.~(\ref{eq:CoherenceBound}), it is immediate to see from its definition (\ref{eq:NAQCDef}) that
\begin{equation}
\label{eq:NAQCMax}
\mathfrak{C}_{\rm{NA}}^{l_1,\rm{re}}(\rho)\leq 3.
\end{equation}
Any NAQC measure satisfies convexity due to the convexity of the underlying coherence measure~\cite{Hu2018}:
\begin{equation}
\mathfrak{C}_{\rm{NA}}\left(\sum_{n}p_n\rho_n\right)\leq \sum_{n}p_n \mathfrak{C}_{\rm{NA}}\left(\rho_n\right).
\end{equation}
This can be proven by noting that, for a convex mixture (\ref{eq:convexsum}), the conditional quantum state $\rho^a_j$ can be expressed as another convex combination of conditional quantum states:
\begin{equation}
\rho^a_j=\sum_{n}\frac{p_np_{n,j}^a}{p^a_j}\rho^a_{n,j}\equiv \sum_n q_{n,j}^a\rho^a_{n,j}, \quad p^a_j=\sum_{n} p_np_{n,j}^a,
\end{equation}
with $p_{n,j}^a,\rho^a_{n,j}$ the corresponding conditional probability and quantum state associated to $\rho_n$. Hence,
\begin{align}
\label{eq:NAQCConvexProof}
\mathfrak{C}_{\rm{NA}}\left(\sum_{n}p_n\rho_n\right)&=\frac{1}{2}\sum_{i\neq j, a}p^a_j\mathfrak{C}_{i}\left(\sum_n q_{n,j}^a\rho^a_{n,j}\right)\\
\nonumber &\leq\frac{1}{2}\sum_{i\neq j, a}\sum_n q_{n,j}^ap^a_j\mathfrak{C}_{i}\left(\rho^a_{n,j}\right)\\
\nonumber &=\frac{1}{2}\sum_{i\neq j, a}\sum_n p_np_{n,j}^a\mathfrak{C}_{i}\left(\rho^a_{n,j}\right)\\
\nonumber &=\sum_{n}p_n \mathfrak{C}_{\rm{NA}}\left(\rho_n\right).
\end{align}
Thus, the absence of NAQC is a convex property (in the same fashion as separability, non-steerability, or Bell locality), since, if all $\rho_n$ satisfy $\mathfrak{C}_{\rm{NA}}\left(\rho_n\right)\leq \bar{\mathfrak{C}}_{\rm{max}}$, then
\begin{equation}
\label{eq:NAQConvex}
\mathfrak{C}_{\rm{NA}}\left(\sum_{n}p_n\rho_n\right)\leq \sum_{n}p_n \mathfrak{C}_{\rm{NA}}\left(\rho_n\right)\leq \sum_{n}p_n\bar{\mathfrak{C}}_{\rm{max}}=\bar{\mathfrak{C}}_{\rm{max}}.
\end{equation}

We remark that the evaluation of NAQC depends on the common orthonormal basis selected for the steering game by Alice and Bob.
Indeed, in more general NAQC scenarios, Alice and Bob can even choose different bases~\cite{Ghosh2023}. 

Using Eqs.~(\ref{eq:DirectionCoherence}),~(\ref{eq:NAQCBloch}), we can explicitly write the LNC measure of NAQC in a given basis as
\begin{align}
\label{eq:NAQCExplicit}
\mathfrak{C}_{\rm{NA}}^{l_1}(\rho)=\frac{1}{4} \sum^3_{j=1}\sum^3_{\substack{i=1 \\ i\neq j}}\sum_{a=\pm 1}|\mathbf{\hat{n}}_i\times\mathbf{B}^+ +a\mathbf{\hat{n}}_i\times(\mathbf{C}\cdot\mathbf{\hat{n}}_j)|.
\end{align}
Expressions simplify considerably for $T$-states~\cite{Horodecki1996} (i.e., unpolarized states, $\mathbf{B}^{\pm}=0$) as
\begin{equation}
\label{eq:ProbabilityTState}
p^{a}_{j}=\frac{1}{2},~\mathbf{B}^+_{a,j}=a\mathbf{C}\cdot\mathbf{\hat{n}}_j.
\end{equation}
In particular, when the correlation matrix is symmetric \footnote{This requirement is added so $\mathbf{C}$ can be diagonalized by applying the same spin rotation in Alice and Bob subspaces. Nevertheless, it can be removed by allowing Alice and Bob to use independent orthogonal bases, which can always chosen so $\mathbf{C}$ is brought to a diagonal form.}, we can characterize NAQC in the orthonormal basis where $\mathbf{C}$ is diagonal with eigenvalues $C_1,C_2,C_3$, so $\mathbf{B}^+_{a,j}=aC_j\mathbf{\hat{n}}_j$. As a result, for $i\neq j$, Eq.~(\ref{eq:DirectionCoherence}) yields 
\begin{align}
\nonumber \mathfrak{C}_{i}^{l_1}(\rho^{a}_{j})&=|\mathbf{\hat{n}}_i\times \mathbf{B}^+_{a,j}|=|C_j|,\\
\mathfrak{C}_{i}^{\rm{re}}(\rho^{a}_{j})&=h\left(\frac{1+\mathbf{\hat{n}}_i\cdot\mathbf{B}^+_{a,j}}{2}\right)-h\left(\frac{1+|\mathbf{B}^+_{a,j}|}{2}\right)\\
\nonumber &=1-h\left(\frac{1+|C_j|}{2}\right).
\end{align}
Hence,
\begin{align}
\label{eq:NAQCTState}
\mathfrak{C}_{\rm{NA}}^{l_1}(\rho)&=\sum^3_{i=1}|C_i|,\\
\nonumber \mathfrak{C}_{\rm{NA}}^{\rm{re}}(\rho)&=3-\sum^3_{i=1}h\left(\frac{1+|C_i|}{2}\right).
\end{align}
Remarkably, the maximum NAQC (\ref{eq:NAQCMax}) is achieved for both LNC and REC measures by a $T$-state with $|C_i|=1$, corresponding to one of the Bell states, in turn also maximally entangled.

The above results provide exact evaluations of NAQC, based on the knowledge of the full density matrix $\rho$. We derive in Appendix~\ref{app:witness} a simple NAQC witness for the LNC measure:
\begin{equation}
\label{eq:NAQCWitness}
W_{\rm{NA}}(\rho)\equiv \sum^3_{i=1}|C_{ii}|,
\end{equation}
satisfying $W_{\rm{NA}}(\rho)\leq \mathfrak{C}_{\rm{NA}}^{l_1}(\rho)$ for an arbitrary quantum state $\rho$. Therefore,
\begin{equation}
W_{\rm{NA}}(\rho)\geq \bar{\mathfrak{C}}^{l_1}_{\rm{max}}=\sqrt{6}
\end{equation}
is always a sufficient condition for a quantum state to be able to manifest NAQC in the corresponding basis where the coefficients $C_{ij}$ are evaluated. Moreover, we show in Eq.~(\ref{eq:WitnessDiagonal}) that $W_{\rm{NA}}(\rho)$ is precisely maximized in the basis where $\mathbf{C}$ is diagonal. 

The above NAQC witness resembles conceptually the Horodecki criterion~\cite{Horodecki1995} for Bell nonlocality, widely used in High-Energy Physics~\cite{Fabbrichesi:2021npl,Afik:2022kwm,Aguilar-Saavedra:2022uye,Fabbrichesi:2024wcd}, which provides a sufficient and necessary condition for a quantum state to be able to manifest nonlocal correlations in a Bell test.

To conclude this section, we note that one can mirror the process so Alice manipulates the coherence of Bob, in whose case all magnitudes are readily evaluated by interchanging $\mathbf{B}^+\leftrightarrow \mathbf{B}^-$, $\mathbf{C}\leftrightarrow \mathbf{C}^{\mathrm{T}}$ in the above expressions. Thus, NAQC is an asymmetric form of quantum correlation, as are quantum discord and steering~\cite{Ollivier2001,Wiseman2007}.

\subsection{$\tau^+\tau^-$ from $e^+e^-$ collisions}\label{subsec:tautheory}

\begin{figure}[h]
\centering
\resizebox{0.45\textwidth}{!}{
\begin{tikzpicture}

 \draw[<-,thick, dashed, line width=3.0] (-0.2,0)--(-5,0) node [left] {\Large $e^-$ beam};
 \draw[<-,thick, dashed, line width=3.0] (0.2,0)--(5,0) node [right] {\Large $e^+$ beam};
 \draw[->,thick, line width=2.0] (0,0)--(5,3) node [right] {\Large $\tau^-$};
 \draw[->,thick, line width=2.0] (0,0)--(-5,-3) node [left] {\Large $\tau^+$};
 \draw[->,thick, draw=orange, line width=2.0] (0,0)--(-1.8,3) node [above, yshift=0.1cm] {\textcolor{orange}{\LARGE\bf$\hat r$}};

 \draw[->,thick, draw=red, line width=2.0] (0,0)--(3,1.8) node [above, yshift=0.2cm] {\textcolor{red}{\LARGE\bf$\hat k$}};

 \draw[->,thick, draw=green, line width=2.0] (0,0)--(3,0) node [above, yshift=0.2cm] {\textcolor{green}{\LARGE\bf$\hat p$}};

\node[blue] at (0,0) {\Huge \bf$\odot$} node [below, yshift=-0.3cm] {\textcolor{blue}{\LARGE\bf$\hat n$}};
\node[blue] at (0,0) {\LARGE \bf$\bullet$};

\draw[thick]
(2,0) arc[start angle=0,end angle=31,radius=2cm];

\node at (1.5,0.45) {\LARGE \bf$\Theta$};

\end{tikzpicture}
}
\caption{Schematic description of the helicity basis.}
\label{fig:helicity}
\end{figure}
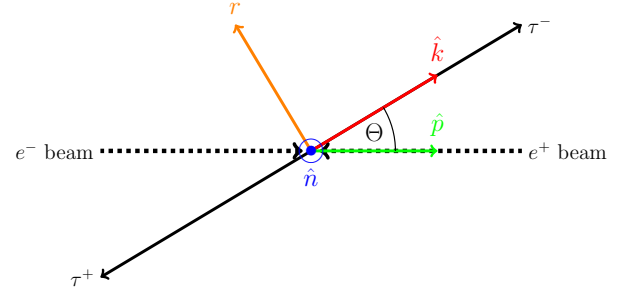

The tau lepton carries spin 1/2 and can therefore be described as a qubit. 
Consequently, a $\tau^+\tau^-$ pair is a two-qubit system, whose density matrix can be written as in Eq.~(\ref{eq:densitymatrix2qubit}), where Alice/Bob correspond to the $\tau^-$/$\tau^+$ lepton, respectively. 
Specifically, the spin quantum state for a given $\tau^+\tau^-$ production process with fixed energy $\sqrt{s}$ and tau-lepton flight direction $\hat{k}$ in the COM frame is described by the density matrix $\rho(\sqrt{s},\hat{k})$.
Here, the indices $i,j$ in Eq.~(\ref{eq:densitymatrix2qubit}) label the directions of a given right-handed orthonormal basis. 
The most common choice is the helicity basis~\cite{Baumgart:2012ay}, given in the $\tau^+\tau^-$ COM frame by the unit vectors $\{\hat{k},\hat{r},\hat{n}\}$. In this work, we take $\hat{r}=(\hat{k}\cos\Theta-\hat{p})/\sin\Theta$, and $\hat{n}=\hat{k}\times \hat{r}$, $\Theta$ being the production angle of the tau lepton with respect to the electron beam direction $\hat{p}$, $\cos\Theta=\hat{k}\cdot\hat{p}$. 
A schematic description of the helicity basis is found in Figure~\ref{fig:helicity}.

In $\tau^+\tau^-$ production through $e^+e^-$ annihilation via electroweak processes, both polarizations are equal, and the correlation matrix is symmetric, with the $n$-axis unpolarized and uncorrelated with respect to the other spin directions:
\begin{equation}
\label{eq:HelicityStructure}
\mathbf{B}^{+}=\mathbf{B}^{-}=\left[\begin{array}{c}
B_{k} \\
B_{r}\\
0 
\end{array}\right],~\mathbf{C}=\mathbf{C}\trans=\left[\begin{array}{ccc}
C_{kk} & C_{kr} & 0 \\
C_{kr} & C_{rr} & 0\\
0 & 0 & C_{nn}
\end{array}\right].
\end{equation}
Remarkably, in contrast to the usual case of QCD processes, such as top-antitop~\cite{Bernreuther1994,Baumgart:2012ay} and bottom-antibottom~\cite{Kats:2023zxb} production at the LHC, we have non-vanishing polarizations even at leading order (LO).
The spin coefficients $B^\pm_{i},C_{ij}$ only depend on the COM energy $\sqrt{s}$ and the production angle $\Theta$; their explicit expressions are specified in Appendix~\ref{app:formulas}. 

Other basis can be defined via rotation from the helicity basis in the COM frame. 
For instance, the beam basis is defined by the unit vectors $\{\hat{x}, \hat{y}, \hat{z}\}$, with $\hat{z}$ along the beam, $\hat{z}=\hat{p}$, while $\hat{x},\hat{y}$ point at transverse directions. 
Importantly, the beam basis points at fixed directions of space, in contrast to the helicity basis, which varies according to the kinematics of the underlying event. 
This difference implies that, after integrating the spin coefficients over phase space, the beam basis leads to \textit{bona-fide} spin quantum states, while event-dependent basis, such as the helicity basis, lead to so-called \textit{fictitious} quantum states~\cite{Afik:2020onf, Afik:2022kwm, Cheng2023, Cheng:2024btk}. 
Nevertheless, fictitious states are a valuable construction to analyze the presence of quantum correlations.

\begin{figure*}[!htbp]
\centering
\includegraphics[width=1\linewidth]{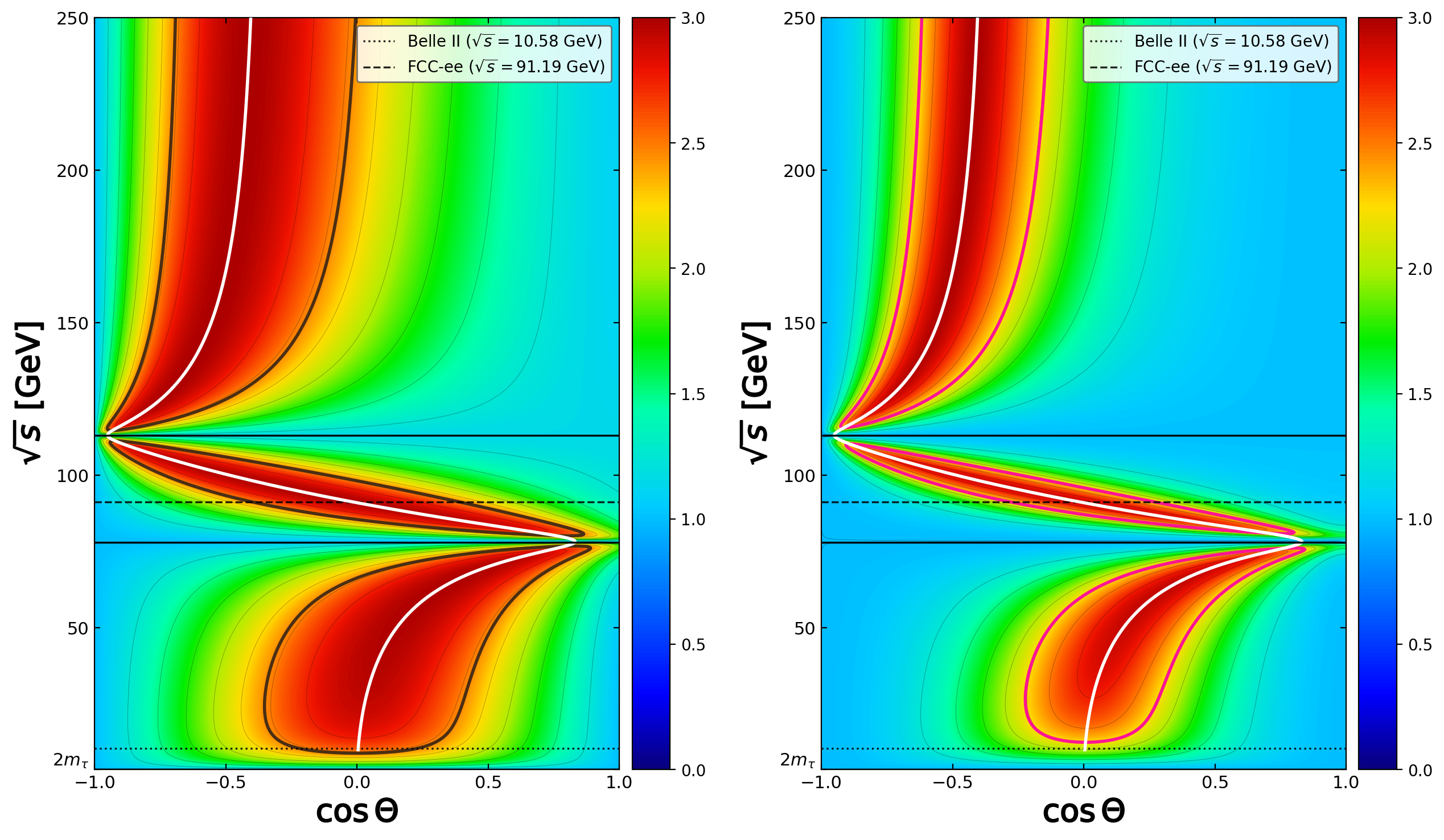}
\caption{$\mathfrak{C}_{\rm{NA}}^{l_1}$ (left) and $\mathfrak{C}_{\rm{NA}}^{\rm{re}}$ (right) in the diagonal basis for $e^+e^- \to \tau^+\tau^-$ production over the $(\cos\Theta,\sqrt{s})$ plane. The brown and magenta contours delimitate the NAQC boundaries, $\mathfrak{C}_{\rm{NA}}^{l_1}=\bar{\mathfrak{C}}^{l_1}_{\rm{max}}=\sqrt{6}$ and $\mathfrak{C}_{\rm{NA}}^{\rm{re}}=\bar{\mathfrak{C}}^{\rm{re}}_{\rm{max}}\approx 2.23$, respectively. Solid white line goes over the center of the NAQC bands, where spin correlations are maximized, while the horizontal solid lines indicate the COM energies $\sqrt{s}_{\pm}$ where transverse spin correlations vanish (see main text). Horizontal dotted and dashed lines mark the COM energies of Belle~II ($\sqrt{s}=10.58$~GeV) and FCC-$ee$ ($\sqrt{s}=91.19$~GeV), respectively.}
\label{fig:naqc_maps}
\end{figure*}

Another useful event-dependent basis is the diagonal basis~\cite{Afik:2022kwm,Cheng2023,Cheng:2024btk} (sometimes referred to as the ``off-diagonal'' basis~\cite{Parke1996}), defined as that diagonalizing the spin-correlation matrix, given by the vectors $\{\hat{u}_{i}\}^3_{i=1}$ with eigenvalues $\{C_{i}\}^3_{i=1}$, $\mathbf{C}\cdot\hat{u}_{i}=C_i\hat{u}_{i}$. We stress that the existence of this common basis for both $\tau^+$ and $\tau^-$ is possible because the correlation matrix is symmetric, and thus it can be diagonalized by the appropriate rotation of the helicity basis; see discussion after Eq.~(\ref{eq:ProbabilityTState}). Specifically, from the structure of Eq.~(\ref{eq:HelicityStructure}), we see that the correlation matrix is diagonalized by a certain rotation in the $\{\hat{k},\hat{r}\}$ plane so the diagonal basis takes the form $\{\hat{u}_{+},\hat{u}_{-},\hat{n}\}$, whose eigenvalues are $\{C_{+},C_{-},C_{nn}\}$, with
\begin{equation}
\label{eq:eigenvalues}
C_{\pm}=\frac{C_{kk}+C_{rr}}{2}\pm\sqrt{\left(\frac{C_{kk}-C_{rr}}{2}\right)^{2}+C_{kr}^{2}}.
\end{equation}
The diagonal basis is particularly advantageous for quantum information purposes, as it optimizes the sensitivity to a number of quantum observables~\cite{Cheng2023,Cheng:2024btk}.

The spin-polarization vectors $\mathbf{B}^{\pm}$ and the spin-correlation matrix $\mathbf{C}$ can be measured from the angular distribution of the $\tau^+\tau^-$ decay products. 
Specifically, the angular distribution of the flight directions $\mathbf{\hat{q}}_{\pm}$ of two given products of the $\tau^{\mp}$ decay in their respective parent rest frames (after boosting from the COM frame) takes the simple form~\cite{Han:2025ewp}
\begin{equation}
\label{eq:TauCrossSectionNormalized}
\begin{split}
\frac{1}{\sigma}\frac{\mathrm{d}\sigma}{\mathrm{d}\Omega_{+}\mathrm{d}\Omega_{-}} =~~~~~~~~~~~~~~~~~~~~~~~~~~~~~~~~~~~~~~~~~~~~~ \\
\frac{1+\alpha^+\mathbf{B}^{+}\cdot\mathbf{\hat{q}}_{+}+\alpha^-\mathbf{B}^{-}\cdot\mathbf{\hat{q}}_{-}+\alpha^+\alpha^-\mathbf{\hat{q}}_{+}\cdot \mathbf{C} \cdot\mathbf{\hat{q}}_{-}}{(4\pi)^2},
\end{split}
\end{equation}
where $\alpha^\pm$ are the so-called spin-analyzing powers (we recall that the vectors $\mathbf{B}^{\pm}$ are the $\tau^{\mp}$ polarizations) and $\Omega_{\pm}$ are the solid angles associated to $\mathbf{\hat{q}}_{\pm}$. 
Thus, by appropriately fitting this expression in the corresponding region of the $\tau^+\tau^-$ phase space, one can retrieve all the spin coefficients $\mathbf{B}^{\pm},\mathbf{C}$, performing in this way the quantum tomography of the $\tau^+\tau^-$ spin quantum state~\cite{Afik:2020onf,Afik:2022kwm}. 
Once the quantum state is known, all QI magnitudes can be readily evaluated.

In particular, tau leptons decay to lighter leptons with neutrinos, namely electrons ($\tau^- \to \nu_\tau e^- \bar\nu_e$) or muons ($\tau^- \to \nu_\tau \mu^- \bar\nu_\mu$), or to hadrons. Among all, the single-prong channel $\tau^- \to \pi^- \nu_\tau $ is the optimal one for polarimetric purposes~\cite{Fabbrichesi:2024wcd} as the pion exhibits nearly maximal spin-analyzing power. State-of-the-art methods for neutrino reconstruction through machine learning~\cite{Zhang:2025mmm} and for quantum tomography~\cite{Ai2026} in $e^+e^-\to\tau^+\tau^- \to \pi^+\bar{\nu}_{\tau}\pi^-\nu_{\tau}$ processes have been recently developed. 

Finally, we note that, while NAQC is in general asymmetric between Alice and Bob, the approximate $CP$-invariance of the Standard Model imposes $\mathbf{B}^{+}=\mathbf{B}^{-}$ and $\mathbf{C}=\mathbf{C}\trans$ [see Eq.~(\ref{eq:HelicityStructure})], so NAQC is symmetric between $\tau^-$ and $\tau^+$. Therefore, our NAQC evaluation can be indistinctively applied to both particles.

\section{Results}\label{sec:Results}

\subsection{NAQC in $e^+e^- \to \tau^+\tau^-$}

We begin by analyzing the presence of NAQC in $\tau^+\tau^-$ pairs produced via $e^+e^-$ annihilation through the LNC and REC measures, retaining the full dependence on the production kinematics. 
Both NAQC measures are analytically evaluated in the diagonal basis using LO electroweak theory; see Eqs.~(\ref{eq:HelicityStructure}),~(\ref{eq:eigenvalues}) and ensuing discussion. We choose the diagonal basis for the NAQC computation because it maximizes the witness (\ref{eq:NAQCWitness}); see also Sec.~\ref{subsec:Perspective}.

Figure~\ref{fig:naqc_maps} shows the resulting landscape of $\mathfrak{C}_{\rm{NA}}^{l_1}$ (left panel) and $\mathfrak{C}_{\rm{NA}}^{\rm{re}}$ (right panel) over the $(\cos\Theta,\sqrt{s})$ plane, where the brown and magenta contours indicate the corresponding NAQC boundaries, see Eq.~(\ref{eq:NAQCBound}).
Three disconnected NAQC bands emerge: i) a low-energy strip, extending from pair threshold up to $\sqrt{s}=\sqrt{s}_-\simeq 78$~GeV; ii) a narrow band around the $Z$ resonance, strongly tilted in $\cos\Theta$ by the $\gamma$--$Z$ interference; iii) a high-energy strip, extending above $\sqrt{s}=\sqrt{s}_+\simeq 113$~GeV.

The structure of the NAQC bands can be easily understood within an ultrarelativistic approximation, which holds in most of phase space due to the smallness of the tau mass (see Appendix~\ref{app:App}). In this limit, Eq.~(\ref{eq:Ultrarelativistic}), the correlation matrix is already diagonal in the helicity basis, satisfying $C_{kk}=1$ and $C_{nn}=-C_{rr}\equiv C$, and only the longitudinal polarization $B^\pm_k$ is non-vanishing. As a result, the amount of NAQC is controlled by the behavior of the transverse correlation coefficient $C$. The NAQC bands are centered precisely around the maxima of $C$ (white curve), given by Eq.~(\ref{eq:MaxCorrelations}). On the other hand, the two gaps at $\sqrt{s}=\sqrt{s}_{\pm}$ (horizontal solid lines) separating the NAQC bands correspond to the COM energies where $C=0$, Eq.~(\ref{eq:MinimalCorrelations}). For comparison, in the ultrarelativistic limit, the condition for Bell nonlocality simply reads $C\neq 0$, demonstrating that NAQC is strictly a more demanding criterion here.

Since the experimental sensitivity is controlled not only by the size of the quantum signal but also by the available number of events, we also evaluate the differential cross section, which at LO takes the form~\cite{Gonzalez-Sprinberg:2000lzf}: 
\begin{equation}
\label{eq:dsigma}
\frac{\mathrm{d}\sigma}{\mathrm{d}\cos\Theta} = K\,\frac{\beta}{s}\,c_0(\sqrt{s},\cos\Theta),
\end{equation}
with $K=0.6785~\mathrm{nb\cdot GeV^2}$ an overall factor, $\beta=\sqrt{1-4m_\tau^2/s}$ the tau COM velocity, $m_\tau$ is the tau mass, and $c_0$ the spin-summed rate function [see Eq.~(\ref{eq:c0})].

\begin{figure}[t]
\centering
\includegraphics[width=1\linewidth]{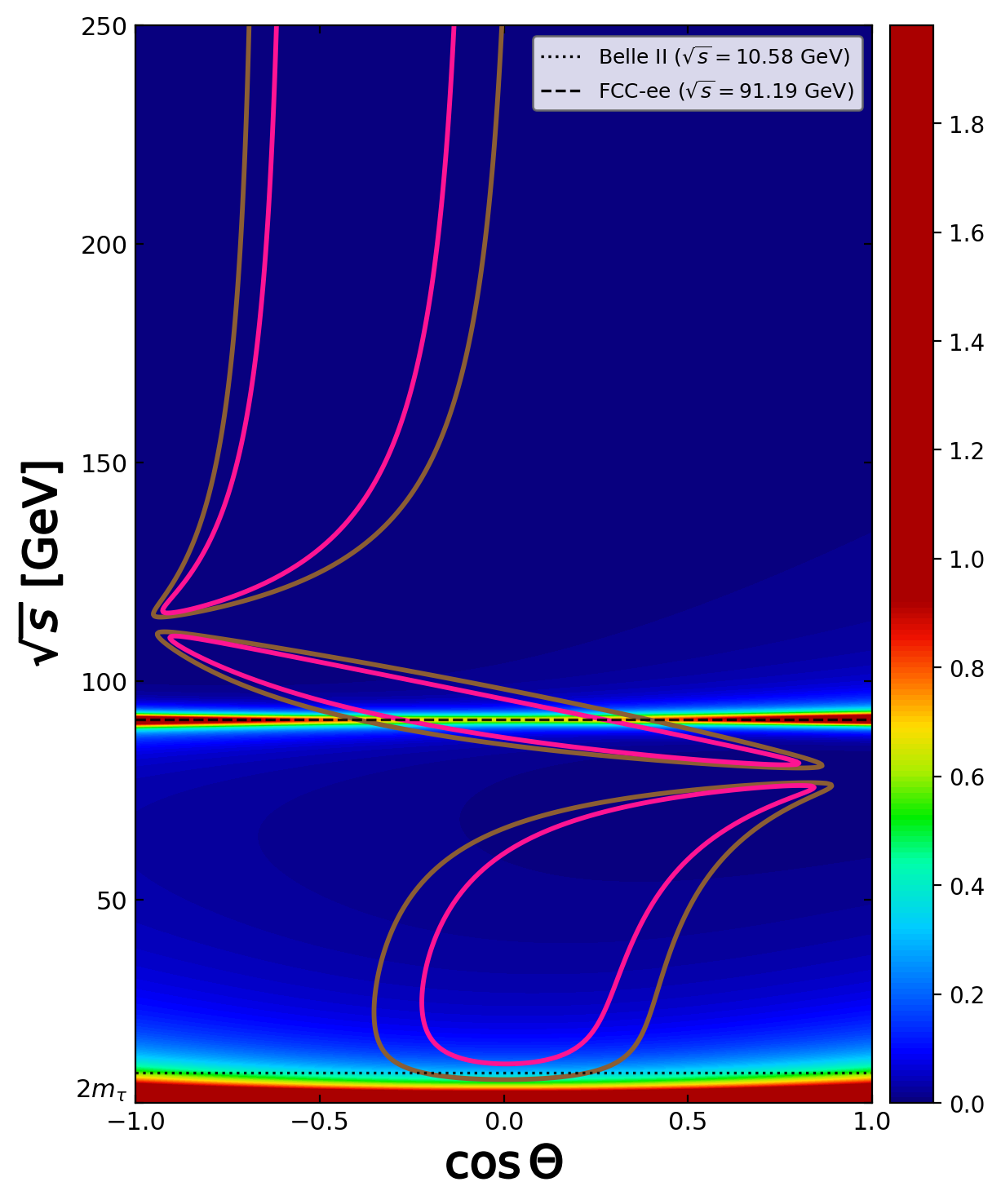}
\caption{Differential cross section in nb for $e^+e^- \to \tau^+\tau^-$ production over the $(\cos\Theta,\sqrt{s})$ plane, including the NAQC boundaries from Figure~\ref{fig:naqc_maps} (brown contour for $\mathfrak{C}_{\rm{NA}}^{l_1}$ and magenta for $\mathfrak{C}_{\rm{NA}}^{re}$). Horizontal dotted and dashed lines mark the COM energies of Belle~II ($\sqrt{s}=10.58$~GeV) and FCC-$ee$ ($\sqrt{s}=91.19$~GeV), respectively.}
\label{fig:naqc_masked}
\end{figure}

Figure~\ref{fig:naqc_masked} displays the differential cross section, along with the NAQC boundaries. The cross-section maximizes at the $Z$ pole ($\sqrt{s}= m_Z \approx 91.19$~GeV), where the production mechanism is dominated by the on-shell $Z$ boson and it is therefore resonantly enhanced, and near the pair-production threshold. 
Accordingly, the main source of $\tau^+\tau^-$ pairs exhibiting NAQC is concentrated in a band around the $Z$ resonance, with a secondary, weaker contribution at low energies.
This analysis then identifies both the low-energy region and the $Z$ pole as the phase-space sectors where NAQC is most favorably accessible in real experiments. 
It should be noted that NAQC is not present if the COM energy is too low, close to the production threshold, $\sqrt s\sim 2m_\tau$. 

Both Figures~\ref{fig:naqc_maps},\ref{fig:naqc_masked} clearly show that the LNC measure reflects NAQC in broader regions of phase space than the REC measure, so hereafter we restrict our analysis to the LNC measure. This observation is in agreement with the results of previous works in the literature~\cite{Ding2019,Rai:2025qke}.

\begin{figure*}[!htbp]
\centering
\includegraphics[width=1\linewidth]{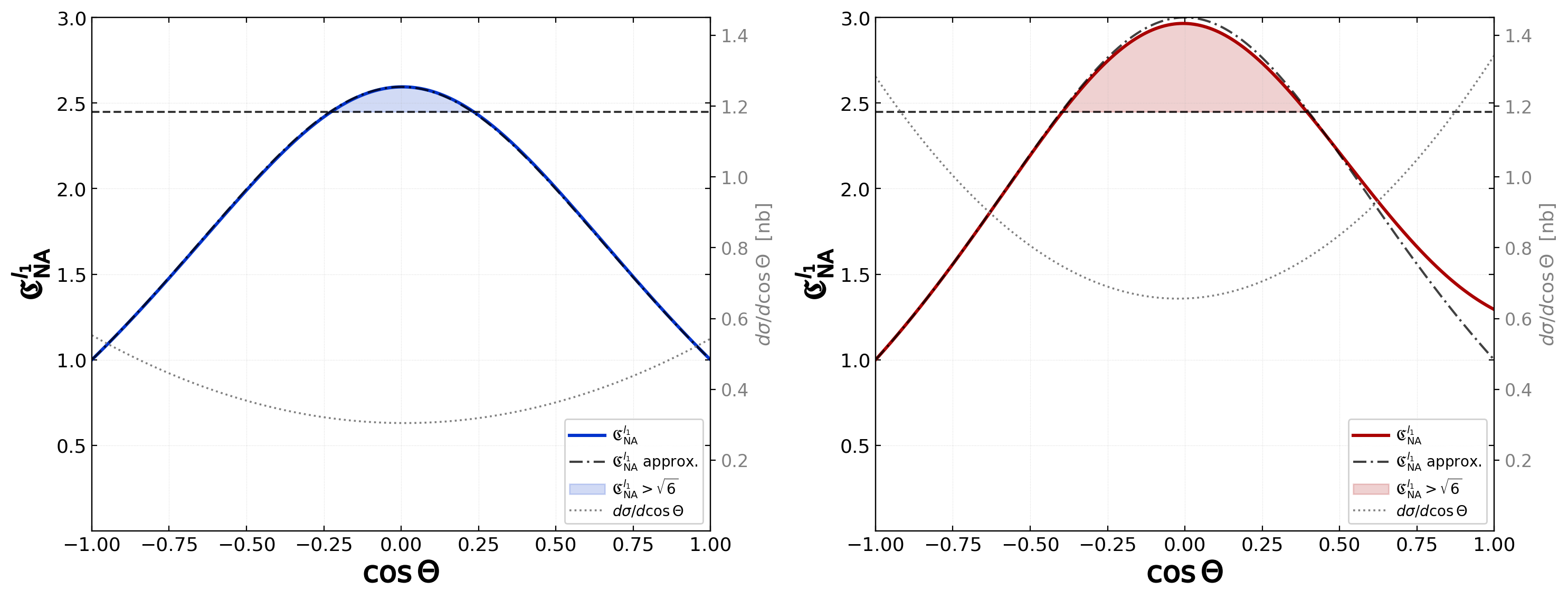}
\caption{$\mathfrak{C}_{\rm{NA}}^{l_1}$ (solid line) in the diagonal basis as a function of $\cos\Theta$ at $\sqrt{s}=10.58$~GeV (Belle~II, left) and $\sqrt{s}=91.19$~GeV (FCC-$ee$, right). The black dashed-dotted line represents the value of $\mathfrak{C}_{\rm{NA}}^{l_1}$ obtained from the approximations in Eqs.~(\ref{eq:SimpleBelle}),~(\ref{eq:SimpleFCC}). The horizontal dashed line is the NAQC boundary $\mathfrak{C}_{\rm{NA}}^{l_1}=\sqrt{6}$, and the shaded area marks the region where NAQC is present. The grey dotted curve is the differential cross section $d\sigma/d\cos\Theta$.}
\label{fig:1dxsec}
\end{figure*}

\subsection{NAQC at Belle~II and FCC-$ee$}

We now focus on the specific cases of $\sqrt{s}=10.58$~GeV ($\beta\approx 0.9419$, horizontal dotted line in Figs.~\ref{fig:naqc_maps},~\ref{fig:naqc_masked}), corresponding to the operation point of Belle and Belle~II, and of the $Z$ pole, $\sqrt{s}=m_Z\approx 91.19$~GeV ($\beta\approx 0.9992$, horizontal dashed line in Figs.~\ref{fig:naqc_maps},~\ref{fig:naqc_masked}), relevant for the Tera-$Z$ run of FCC-$ee$.
Belle and Belle~II operate at the same COM energy and thus probe the same underlying physics processes. We refer primarily to Belle~II throughout the text, given its substantially larger expected dataset and upgraded detector capabilities.

Figure~\ref{fig:1dxsec} shows $\mathfrak{C}_{\rm{NA}}^{l_1}$ (solid color line) together with the differential cross section (dotted gray line) for Belle~II (left panel) and FCC-$ee$ (right panel) as a function of $\cos\Theta$. 
In both cases, we observe that NAQC is present in a central window, peaked around $\cos\Theta=0$. 
However, the differential cross section also reaches a minimum at $\cos\Theta=0$, so the number of events is anti-correlated with the NAQC signal. We can understand these results using simple approximations; see Appendix~\ref{app:App} for the details.

For $\sqrt{s}\ll m_Z$, like in Belle~II, the photon channel dominates over the $Z$ channel. As a result, the spin quantum state is a $T$-state determined by [see Eq.~(\ref{eq:FormFactorsBelle}) and ensuing discussion]:
\begin{align}
\label{eq:SimpleBelle}
c_0&\simeq 48(2-\beta^2\sin^2\Theta),\\
\nonumber C_{kk}&\simeq \frac{2-(2-\beta^2)\sin^2\Theta}{2-\beta^2\sin^2\Theta},\\
\nonumber C_{rr}&\simeq\frac{(2-\beta^2)\sin^2\Theta}{2-\beta^2\sin^2\Theta},\\
\nonumber C_{nn}&\simeq-\frac{\beta^2\sin^2\Theta}{2-\beta^2\sin^2\Theta},\\
\nonumber C_{kr}&\simeq-\,\sqrt{1-\beta^2}\;\frac{2\cos\Theta\sin\Theta}{2-\beta^2\sin^2\Theta}.
\end{align}
Remarkably, this is the same spin quantum state as that of top-antitop and bottom-antibottom quarks produced from quark-antiquark annihilation~\cite{Afik:2020onf,Afik:2022kwm,Afik:2025grr}.

Regarding NAQC, since it is a $T$-state and we focus on the diagonal basis, the former can be evaluated using Eq.~(\ref{eq:NAQCTState}). In particular, $C_{+}=1>C_{-}=-C_{nn}>0$, so
\begin{equation}
|C_+|+|C_-|=C_++C_-=C_{kk}+C_{rr},
\end{equation}
and then
\begin{equation}
\label{eq:NAQCBelle}
\mathfrak{C}^{l_1}_{\rm{NA}}=C_{kk}+C_{rr}+|C_{nn}|=\frac{2+\beta^2\sin^2\Theta}{2-\beta^2\sin^2\Theta}.
\end{equation}
This is an even function in $\cos\Theta$, which has a maximum at $\cos\Theta=0$,
\begin{equation}
\mathfrak{C}^{l_1}_{\rm{NA}}(\sqrt{s},\cos\Theta=0)=\frac{2+\beta^2}{2-\beta^2}.
\end{equation} 
Hence, in order to have NAQC ($\mathfrak{C}^{l_1}_{\rm{NA}}>\sqrt{6}$),
\begin{equation}
\label{eq:CriticalBeta}
\beta>\beta_c=\sqrt{\frac{14-4\sqrt{6}}{5}}\approx 0.917.
\end{equation}
Equivalently, the COM energy must be above the critical value $\sqrt{s_c}$,
\begin{equation}
\label{eq:CriticalCOM}
\sqrt{s}>\sqrt{s_c}=2m_\tau\sqrt{\frac{4\sqrt{6}}{3}+3}\approx 8.9~\textrm{GeV}.
\end{equation}
This condition automatically rules out $e^+e^-$ experiments with a COM energy lower than 8.9~GeV for our proposed observation of NAQC, such as BESIII, while allowing it at Belle and Belle~II. 
On the other hand, notice that the quantum state (\ref{eq:SimpleBelle}) displays Bell nonlocality for any value $\beta\sin\Theta\neq 0$~\cite{Afik:2020onf,Afik:2022kwm,Afik:2025grr}. 
This shows again that NAQC is strictly a tighter form of quantum correlation than Bell nonlocality here. 
Interestingly, $\mathfrak{C}^{l_1}_{\rm{NA}}$ takes the same expression of an entanglement witness in Eq.~(\ref{eq:NAQCBelle}), with $\mathfrak{C}^{l_1}_{\rm{NA}}>1$ implying entanglement~\cite{Afik:2020onf,Afik:2022kwm,Afik:2025grr}.

For COM energies above the critical threshold, $\sqrt{s}>\sqrt{s}_c$, NAQC is expected in the angular window
\begin{equation}
\label{eq:NAQCWindowBelle}
|\cos\Theta|<\sqrt{1-\frac{\beta^2_c}{\beta^2}}.
\end{equation}
Physically, NAQC emerges in this limit because, for transverse production ($\cos\Theta=0$), the $\tau^+\tau^-$ system approaches in the ultrarelativistic limit ($\beta=1$) a spin-triplet with zero projection along the $n$-axis~\cite{Afik:2020onf,Afik:2022kwm,Afik:2025grr}:
\begin{align}
\label{eq:triplet}
C_{ij}&=\delta_{ij}-2\hat{n}_i\hat{n}_j,\\
\nonumber \rho&=\ket{\Psi_{\hat{n}}^{+}}\bra{\Psi_{\hat{n}}^{+}},~\ket{\Psi_{\hat{n}}^{+}}=\frac{\ket{+\hat{n},-\hat{n}}+\ket{-\hat{n},+\hat{n}}}{\sqrt{2}}.
\end{align}
This is indeed one of the Bell states, which displays maximal NAQC, as explained after Eq.~(\ref{eq:NAQCTState}).

On the other hand, in the $Z$-dominated limit of FCC-$ee$, $\sqrt{s}=m_Z$, we can work directly in the ultrarelativistic limit. For simplicity, we neglect the polarization and approximate the quantum state as a $T$-state of the form
\begin{align}
\label{eq:SimpleFCC}
c_0&\simeq 8052(1+\cos^2\Theta),\\
\nonumber C_{kk}&\simeq 1,\\
\nonumber C_{nn}&\simeq -C_{rr}\simeq C\simeq \frac{\sin^2\Theta}{1+\cos^2\Theta}.
\end{align} 
Although a more precise approximation accounting for the non-vanishing polarization can be made [see Eq.~(\ref{eq:UltrarelativisticFCC})], the advantage of this model is that, apart from being accurate yet simple, it does not depend on the electroweak couplings, allowing for straightforward evaluations.

The resulting NAQC reads
\begin{equation}
\mathfrak{C}^{l_1}_{\rm{NA}}=\frac{3-\cos^2\Theta}{1+\cos^2\Theta},
\end{equation}
which predicts NAQC in the angular window
\begin{equation}
\label{eq:NAQCWindowFCC}
|\cos\Theta|<\sqrt{\frac{4\sqrt{6}-9}{5}}\simeq 0.40.
\end{equation}
Similarly to the photon case, the $\tau^+\tau^-$ system displays maximal NAQC at $\cos\Theta=0$, where it approaches a spin-triplet with zero projection, now along the $r$-axis. On the other hand, for both photon and $Z$-boson production mechanisms, the cross section is a parabola in $\cos\Theta$, reaching its minimum at $\cos\Theta=0$.

\subsection{Angular average}
\label{subsec:angular_average}

In actual experiments, magnitudes are necessarily averaged over phase space due to the finite size of the analyzed regions. In the specific case of $e^+e^- \to \tau^+\tau^-$ production, where the COM energy $\sqrt{s}$ is fixed by the colliding beam, the average is performed over a subset $\Sigma$ of the sphere, $\Sigma \subseteq S^2$, determined by the imposed requirements on the tau direction $\hat{k}$. The expectation value of the observable $X$ in the region $\Sigma$ is then given by the angular average 
\begin{align}
\label{eq:AngularAverage}
\nonumber X_{\Sigma}(\sqrt{s})&=\frac{1}{\sigma_\Sigma}\int_{\Sigma}\mathrm{d}\Omega\,\frac{\mathrm{d}\sigma}{\mathrm{d}\Omega}X,\\
\sigma_\Sigma(\sqrt{s})&=\int_{\Sigma}\mathrm{d}\Omega\,\frac{\mathrm{d}\sigma}{\mathrm{d}\Omega},
\end{align}
where $\Omega$ is the solid angle associated to $\hat{k}$ and $\sigma_\Sigma$ is the total cross section in the region $\Sigma$. 

The above reasoning also applies to the spin quantum state, leading to the density matrix~\cite{Afik:2020onf,Afik:2022kwm} 
\begin{equation}
\label{eq:AngularQuantumState}
\rho_{\Sigma}(\sqrt{s})=\frac{1}{\sigma_\Sigma(\sqrt{s})}\int_{\Sigma}\mathrm{d}\Omega\,\frac{\mathrm{d}\sigma}{\mathrm{d}\Omega}\rho(\sqrt{s},\hat{k}),
\end{equation}
whose spin coefficients are in turn the angular average (\ref{eq:AngularAverage}) of the spin coefficients of the individual states $\rho(\sqrt{s},\hat{k})$. 
In this way, $\rho_\Sigma$ describes the spin quantum state of the $\tau^+\tau^-$ system in the region $\Sigma$. We note that, when the individual density matrices $\rho(\sqrt{s},\hat{k})$ in the r.h.s. of Eq.~(\ref{eq:AngularQuantumState}) are evaluated in the helicity or diagonal basis, $\rho_\Sigma$ describes a fictitious state~\cite{Afik:2022kwm}. 
However, since the set of states which do not exhibit NAQC is convex, Eq.~(\ref{eq:NAQConvex}), the presence of NAQC in $\rho_{\Sigma}$ necessarily implies that some of the individual states $\rho(\sqrt{s},\hat{k})$ must also display NAQC. 
Thus, we can use the state $\rho_\Sigma$, whose quantum tomography can be directly performed by fitting its spin coefficients from Eq.~(\ref{eq:TauCrossSectionNormalized}), to experimentally signal the presence of NAQC. 

In particular, since the relevant magnitudes are (approximately) even functions of $\cos\Theta=0$ (see Figure~\ref{fig:1dxsec}), we will evaluate the expectation values in the region $\Sigma$ defined by $|\cos\Theta|<x_c$. Moreover, since the density-matrix elements only depend on $x=\cos\Theta$ for the helicity and diagonal bases, $\rho(\sqrt{s},\hat{k})=\rho(\sqrt{s},x)$, Eq.~(\ref{eq:AngularQuantumState}) is reduced to
\begin{align}
\label{eq:PolarQuantumState}
\rho_{\Sigma}(\sqrt{s},x_c)&=\frac{1}{\sigma_\Sigma(\sqrt{s},x_c)}\int^{x_c}_{-x_{c}}\mathrm{d}x\,\frac{\mathrm{d}\sigma}{\mathrm{d}x}\rho(\sqrt{s},x),\\
\nonumber \sigma_\Sigma(\sqrt{s},x_c)&=\int^{x_c}_{-x_{c}}\mathrm{d}x\,\frac{\mathrm{d}\sigma}{\mathrm{d}x}.
\end{align}
Notice that, when $x_c=1$, we do not impose any angular requirement, averaging over all possible directions. In particular, 
\begin{equation}
\sigma(\sqrt{s})=\sigma_\Sigma(\sqrt{s},1)
\end{equation}
is the total cross section for a given COM energy $\sqrt s$. The NAQC for the averaged density matrix is then
\begin{equation}
\label{eq:AngularAverageNAQC}
\mathfrak{C}^{l_1}_{\rm{NA}}(\sqrt{s},x_c)\equiv \mathfrak{C}^{l_1}_{\rm{NA}}\left[ \rho_{\Sigma}(\sqrt{s},x_c)\right].
\end{equation}
For practical purposes, one can use the simple witness of Eq.~(\ref{eq:NAQCWitness}) to certify the presence of NAQC:
\begin{equation}
\label{eq:AngularAverageNAQCWitness}
W_{\rm{NA}}(\sqrt{s},x_c)\equiv W_{\rm{NA}}\left[ \rho_{\Sigma}(\sqrt{s},x_c)\right].
\end{equation}
Nevertheless, both magnitudes give essentially the same results. Indeed, using the simple approximations of Eqs.~(\ref{eq:SimpleBelle}),~(\ref{eq:SimpleFCC}), we find 
\begin{align}
\nonumber \mathfrak{C}^{l_1}_{\rm{NA}}(\sqrt{s},x_c)&= W_{\rm{NA}}(\sqrt{s},x_c)\simeq\frac{2+\beta^2-\beta^2\dfrac{x^2_c}{3}}{2-\beta^2+\beta^2\dfrac{x^2_c}{3}},\\
\sigma_\Sigma(\sqrt{s},x_c)&\simeq 96K\frac{\beta}{s}x_c\left(2-\beta^2+\beta^2\frac{x^2_c}{3}\right),
\end{align}
for Belle~II and
\begin{align}
\nonumber \mathfrak{C}^{l_1}_{\rm{NA}}(\sqrt{s}=m_Z,x_c)&= W_{\rm{NA}}(\sqrt{s}=m_Z,x_c)\simeq \frac{3-\dfrac{x^2_c}{3}}{1+\dfrac{x^2_c}{3}},\\
\sigma_\Sigma(\sqrt{s}=m_Z,x_c)&\simeq 16104\,K\frac{\beta}{s}x_c\left(1+\frac{x^2_c}{3}\right),
\end{align}
for FCC-$ee$.

\section{Experimental NAQC detection}\label{sec:Experimental}

\subsection{Experimental analysis}\label{subsec:Proposal}

\begin{table}[t]
\centering
\begin{tabular}{lcccc}
\hline\hline

 & $\mathfrak{C}^{l_1}_{\rm{NA},max}$ & $x_{\rm{NAQC}}$ & $f_{\rm{NAQC}}$\\
\hline
 Belle~II & 2.59 & 0.23 & 0.18 \\
FCC-$ee$ & 2.96 & 0.39 & 0.31 \\
\hline\hline
\end{tabular}
\caption{Figures of merit $\mathfrak{C}^{l_1}_{\rm{NA},max},x_{\rm{NAQC}},f_{\rm{NAQC}}$ measuring the maximum, the width, and the fraction of events of the NAQC peak in Figure~\ref{fig:1dxsec}, respectively, for both Belle~II and FCC-$ee$.}
\label{tab:1dresults}
\end{table}

We now address the experimental characterization of NAQC in both Belle~II and FCC-$ee$. As figure of merits, we use the maximum value of the NAQC measure, $\mathfrak{C}^{l_1}_{\rm{NA},max}$, the maximum value of $\cos\Theta$ exhibiting NAQC, $x_{\rm{NAQC}}$, and the fraction of the total number of events present in the region $\Sigma$ exhibiting NAQC, $f_{\rm{NAQC}}$. Due to the even symmetry of $\mathfrak{C}^{l_1}_{\rm{NA}}$ with respect to $\cos\Theta$, $f_{\rm{NAQC}}$ can be computed in terms of $x_{\rm{NAQC}}$ as
\begin{equation}
\label{eq:FNAQC}
f_{\rm{NAQC}}=\frac{\sigma_{\Sigma}(\sqrt{s},x_{\rm{NAQC}})}{\sigma(\sqrt{s})}.
\end{equation}
The results are shown in Table~\ref{tab:1dresults}. In particular, we find $f_{\rm{NAQC}}=0.18$ and $f_{\rm{NAQC}}=0.30$ for Belle~II and FCC-$ee$, respectively, which clearly demonstrates the promising character of both setups for the observation of NAQC.

These results are in very good agreement with those predicted by the simple approximations of Eqs.~(\ref{eq:SimpleBelle}),~(\ref{eq:SimpleFCC}), which for Belle~II yield:
\begin{align}
\nonumber \mathfrak{C}^{l_1}_{\rm{NA},max}&=\frac{2+\beta^2}{2-\beta^2}\approx 2.59,\\
x_{\rm{NAQC}}&=\sqrt{1-\frac{\beta^2_c}{\beta^2}}\approx 0.23,\\
\nonumber f_{\rm{NAQC}}&=x_{\rm{NAQC}}\frac{2-\beta^2+\beta^2\dfrac{x^2_{\rm{NAQC}}}{3}}{2-2\dfrac{\beta^2}{3}}\approx 0.18,
\end{align}
while for FCC-$ee$:
\begin{align}
\nonumber \mathfrak{C}^{l_1}_{\rm{NA},max}&=3,\\
x_{\rm{NAQC}}&=\sqrt{1-\beta^2_c}\approx 0.40,\\
\nonumber f_{\rm{NAQC}}&=x_{\rm{NAQC}}\frac{3+x^2_{\rm{NAQC}}}{4}\approx 0.32.
\end{align}

\begin{figure}[t]
\centering
\includegraphics[width=\linewidth]{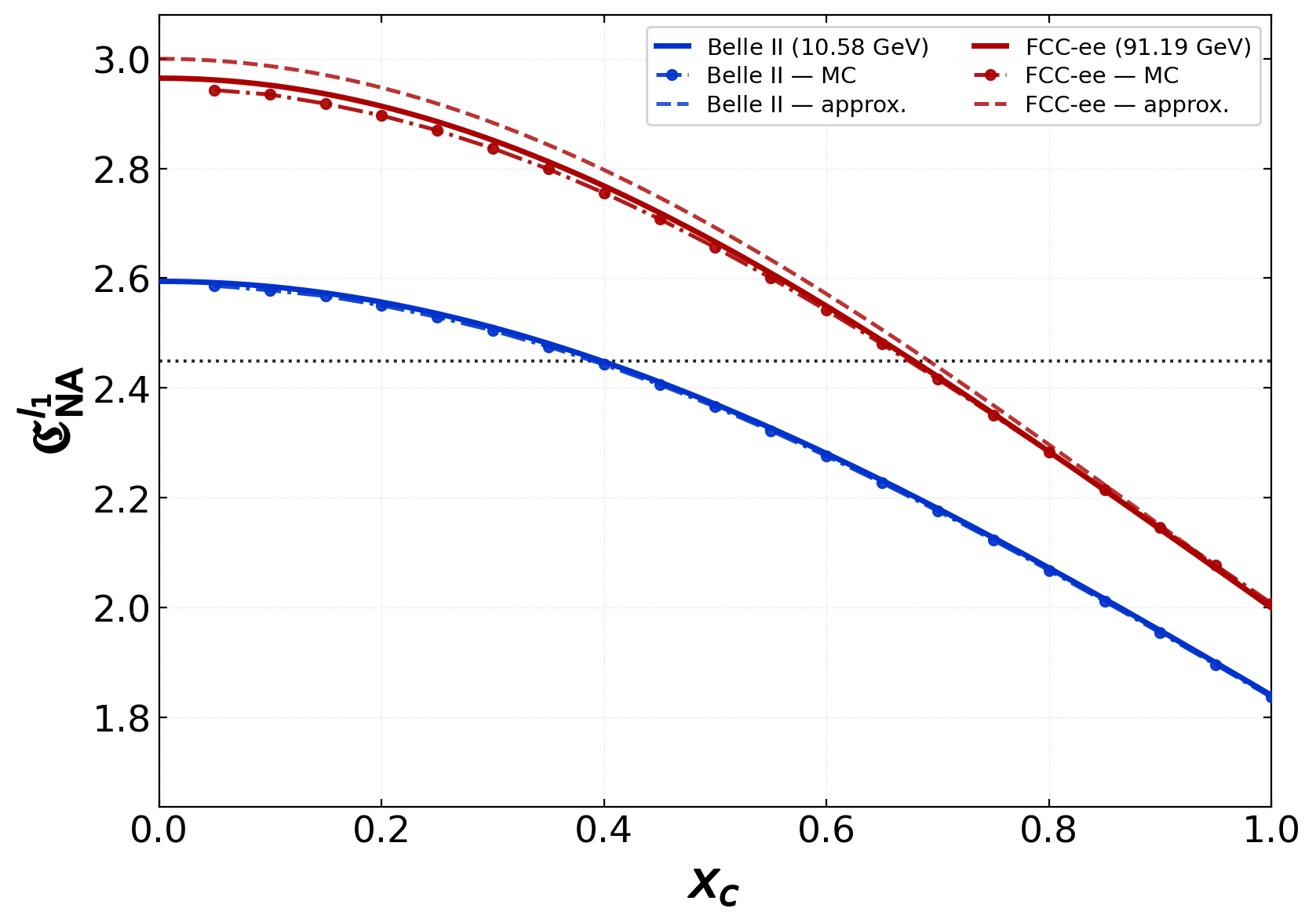}
\caption{NAQC measure of the angular averaged quantum state $\rho_{\Sigma}$ over the selection window $|\cos\Theta|<x_c$, $\mathfrak{C}^{l_1}_{\rm{NA}}(\sqrt{s},x_c)$, for Belle~II (blue) and FCC-$ee$ (red). 
Solid lines show the exact analytical calculation, dashed-dotted lines the results from the MC simulation, and dashed lines the corresponding approximations from Eqs.~(\ref{eq:SimpleBelle}),~(\ref{eq:SimpleFCC}).
The horizontal dotted line indicates the NAQC boundary (\ref{eq:xmax}).}
\label{fig:angcut}
\end{figure}

As explained in Section~\ref{subsec:angular_average}, the actual experimental signal is given in terms of the density matrix $\rho_\Sigma$. 
Specifically, we compute $\mathfrak{C}^{l_1}_{\rm{NA}}(\sqrt{s},x_c)$, Eq.~(\ref{eq:AngularAverageNAQC}), which is shown in Figure~\ref{fig:angcut} for both Belle~II and FCC-$ee$. 
The maximum angular cut $x_{\rm{max}}$ satisfying
\begin{equation}
\label{eq:xmax}
\mathfrak{C}^{l_1}_{\rm{NA}}(\sqrt{s},x_{\rm{max}})=\sqrt{6},
\end{equation}
defines the loosest selection for which $\rho_\Sigma$ retains NAQC. 
We find $x_{\rm{max}}\approx 0.40$ for Belle~II and $x_{\rm{max}}\approx 0.68$ for FCC-$ee$, in great agreement with the theoretical prediction from our simplified model, $x_{\rm{max}}=\sqrt{3}x_{\rm{NAQC}}$, which yields $x_{\rm{max}}\approx 0.40$ and $x_{\rm{max}}\approx 0.69$, respectively. 

\begin{figure*}[!tbp]
\centering
\includegraphics[width=1\linewidth]{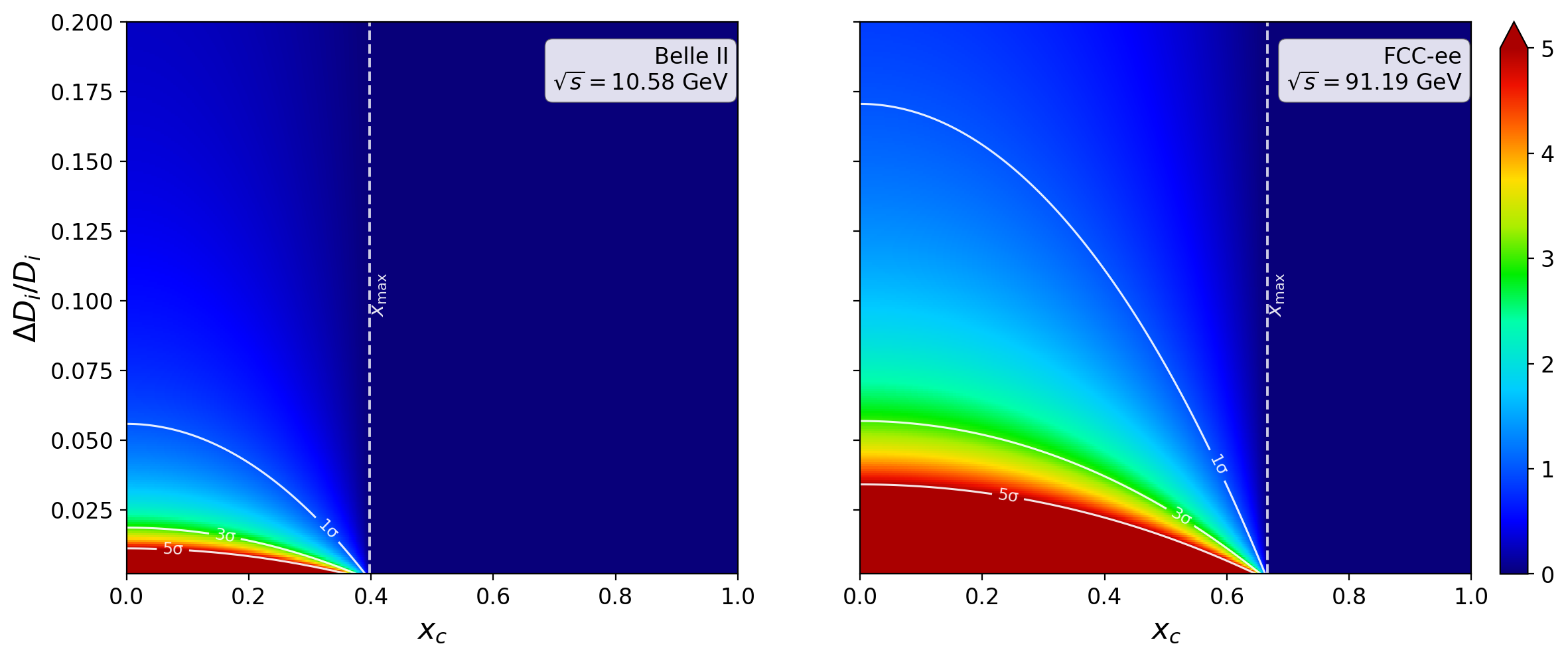}
\caption{Expected significance $\mathcal{S}$ for an NAQC observation in $e^+e^-\to\tau^+\tau^-$, Eq.~\eqref{eq:significance}, as a function of the central angular cut $|\cos\Theta|<x_c$ and the relative uncertainty $\Delta D_i/D_i$. White contours mark $1,3,5\sigma$. Vertical dashed line is the angular cut $x_{\rm{max}}$, Eq.~(\ref{eq:xmax}). Left: Belle~II, $D_i=D_n$, Eq.~(\ref{eq:NAQCWitnessBelleII}). Right: FCC-$ee$, $D_i=D_r$, Eq.~(\ref{eq:NAQCWitnessFCC}).} 
\label{fig:significance}
\end{figure*}

We validate our analytical results using an independent Monte Carlo (MC) simulation. 
The process $e^+e^- \to \tau^+\tau^-$ is generated with {\sc MadGraph}~v3.5.7~\cite{Alwall:2014hca}, and the subsequent $\tau$ decays are simulated with the {\sc TauDecay}~\cite{Hagiwara:2012vz} library, yielding the complete process $e^+e^- \to \tau^+\tau^- \to \pi^+\bar{\nu}_{\tau}\pi^-\nu_{\tau}$. 
No additional cuts are applied beyond the default generator settings. 
The results from the MC simulation are also shown in Figure~\ref{fig:angcut}, demonstrating good agreement between the two approaches.

\subsection{Experimental perspective}\label{subsec:Perspective}

We present an experimental perspective for the observation of NAQC in both Belle~II and FCC-$ee$. For simplicity, we restrict to $\tau^+\tau^- \to \pi^+\bar{\nu}_{\tau}\pi^-\nu_{\tau}$ decays, since pions $\pi^\mp$ possess nearly maximal spin analyzing power, $\alpha^\pm\simeq \pm 1$. We propose
\begin{equation}
D_i=\frac{\textrm{Tr}[P_i\mathbf{C}]}{3}=\frac{-C_{ii}+\sum_{j\neq i}C_{jj}}{3}=\frac{\textrm{Tr}[\mathbf{C}]-2C_{ii}}{3}
\end{equation}
as the observable of interest, where $P_i$ is the matrix implementing a parity transformation along axis $i$. This observable can be extracted from the differential cross section [derived after angular integration from Eq.~(\ref{eq:TauCrossSectionNormalized})]
\begin{equation}
\frac{1}{\sigma}\frac{d\sigma}{d\cos\tilde{\varphi}}=\frac{1-D_i\cos\tilde{\varphi}}{2},~\quad\cos\tilde{\varphi}=\mathbf{\hat{q}}_{+}\cdot P_i \cdot \mathbf{\hat{q}}_{-}.
\end{equation}
We note that $\cos\tilde{\varphi}$ is analogous to the opening angle between the pions, $\cos\varphi=\mathbf{\hat{q}}_{+}\cdot \mathbf{\hat{q}}_{-}$, but with an inverted sign for the $i$-component of one of the pions. Such an observable was originally proposed for measuring top-antitop quark entanglement in the ultrarelativistic regime in Ref.~\cite{Aguilar-Saavedra:2022uye}, and was later experimentally implemented in the entanglement observation by the CMS collaboration of Ref.~\cite{CMS:2024zkc}. 

Remarkably, this observable also provides an experimental NAQC witness [see Eq.~(\ref{eq:NAQCWitness})], since it is immediate to notice that
\begin{equation}
D_i\leq \frac{W_{\rm{NA}}}{3}.
\end{equation}
Hence,
\begin{equation}
\label{eq:ExperimentalNAQCWitness}
D_i\geq \sqrt{\frac{2}{3}}
\end{equation}
is a sufficient condition for the quantum state to exhibit NAQC in the considered orthonormal basis. In addition, measuring this condition automatically certifies the presence of NAQC in the diagonal basis, since the latter maximizes $W_{\rm{NA}}$, Eq.~(\ref{eq:WitnessDiagonal}). 

Specifically, we propose to measure
\begin{equation}
\label{eq:NAQCWitnessBelleII}
D_n=\frac{-C_{nn}+C_{kk}+C_{rr}}{3}
\end{equation}
in Belle~II and
\begin{equation}
\label{eq:NAQCWitnessFCC}
D_r=\frac{-C_{rr}+C_{kk}+C_{nn}}{3}
\end{equation}
in FCC-$ee$, after applying a kinematic cut in the production angle $|\cos\Theta|<x_c$. 
In both cases, based on the signs of the correlation coefficients from Eqs.~(\ref{eq:SimpleBelle}),~(\ref{eq:SimpleFCC}), we have $D_i=W_{\rm{NA}}/3$. 
Furthermore, since $W_{\rm{NA}}\simeq \mathfrak{C}^{l_1}_{\rm{NA}}$ [see discussion around Eq.~(\ref{eq:AngularAverageNAQCWitness})], the proposed witnesses are expected to capture quite accurately the underlying NAQC.

Notice that the proposed measurement is performed in the more conventional helicity basis instead of the targeted diagonal basis. This is because i) for Belle~II, $W_{\rm{NA}}$ takes the same value in both the helicity and diagonal bases [see Eq.~(\ref{eq:NAQCBelle})]; and ii) for FCC-$ee$, the helicity and diagonal bases are equivalent to a very good approximation.
Thus, the witnesses (\ref{eq:NAQCWitnessBelleII}), (\ref{eq:NAQCWitnessFCC}) are also optimized in the helicity basis.

In order to evaluate the expected significance of an NAQC observation, the null hypothesis is defined to be the NAQC limit $D_{i}=\sqrt{2/3}$. 
We then compute 
\begin{equation}
\label{eq:significance}
\mathcal{S}\equiv\max\!\left[\frac{D_{i}-\sqrt{\dfrac{2}{3}}}{\Delta D_{i} },\,0\right]
\end{equation}
as a function of the angular cut, $x_c$, and the relative uncertainty,
$\Delta D_{i}/D_{i}$, where $D_{i}$ is the expected value [theoretically calculated from Eq.~(\ref{eq:AngularAverageNAQCWitness})] and $\Delta D_{i}$ is the uncertainty of the measurement. The significance $\mathcal{S}$ therefore represents the number of measurement uncertainties differing between the expected measurement and the null hypothesis.
Figure~\ref{fig:significance} shows $\mathcal{S}$ over the $(x_c,\Delta D_{i}/D_{i})$ plane for both Belle~II (left panel) and FCC-$ee$ (right panel), with white contours at $1,3,5\sigma$ and the vertical dashed line marking the cut at which $D_{i}=\sqrt{2/3}$ [accurately given by $x_{\rm{max}}$, Eq.~(\ref{eq:xmax})]. 

Of course, one can always reconstruct the full density matrix $\rho_{\Sigma}(\sqrt{s},x_c)$ through quantum tomography after applying an angular cut $|\cos\Theta|<x_c$, and then evaluate the full NAQC measure $\mathfrak{C}^{l_1}_{\rm{NA}}$. However, this requires the measurement of $15$ parameters and the estimation of the associated $15$ uncertainties, while our experimental proposal can signal the presence of NAQC with the measurement of a single parameter, in the same fashion as the entanglement witnesses used in Refs.~\cite{ATLAS:2023fsd,CMS:2024pts,CMS:2024zkc}.

To further assess the prospects for measuring NAQC at $e^+e^-$ colliders, it is useful to estimate the expected number of $e^+e^- \to \tau^+\tau^- \to \pi^+\bar{\nu}_{\tau}\pi^-\nu_{\tau}$ events contained in the regions exhibiting NAQC. We define this number as
\begin{equation}
N_{NAQC} = \sigma(\sqrt{s}) \cdot \mathcal{L}_{int} \cdot \left[BR(\tau^-\to\pi^-\nu_{\tau})\right]^2 \cdot f_{NAQC},
\end{equation}
where $\mathcal{L}_{int}$ is the integrated luminosity and $BR(\tau^-\to\pi^-\nu_{\tau})\approx 0.108$~\cite{ParticleDataGroup:2026mpi} is the branching ratio for $\tau^-\to\pi^-\nu_{\tau}$.
In principle, other $\tau$-lepton decay channels could also be considered to increase the available statistics. For simplicity, however, we restrict the present discussion to the $\tau^+\tau^- \to \pi^+\bar{\nu}_{\tau}\pi^-\nu_{\tau}$ decay channel.
No detector acceptance or reconstruction efficiencies are included in this estimate, as their evaluation requires a dedicated detector-level analysis beyond the scope of this work.

The COM energy at Belle and Belle~II is predominantly $\sqrt{s}=10.58$~GeV. 
Belle collected approximately $1$~ab$^{-1}$ of integrated luminosity, most of it at this energy~\cite{Belle:2012iwr}, while Belle~II is expected to collect up to $50$~ab$^{-1}$~\cite{Belle-II:2018jsg}. 
At the FCC-$ee$, an integrated luminosity of approximately $125$~ab$^{-1}$ is expected at $\sqrt{s}=91.19$~GeV~\cite{Benedikt:2025kwi}. 
The LEP experiments operated at COM energies up to $209$~GeV. 
As an example, the DELPHI collaboration recorded $116$~pb$^{-1}$ at the $Z$ pole~\cite{DELPHI:2000wje}.
Table~\ref{tab:nevents} summarizes the estimated number of $\tau^+\tau^- \to \pi^+\bar{\nu}_{\tau}\pi^-\nu_{\tau}$ events expected to exhibit NAQC in each of these datasets, given by $N_{NAQC}$.
Among the considered facilities, the FCC-$ee$ offers the largest statistical sample and therefore the strongest prospects for observing NAQC. Belle and Belle~II also provide substantial event samples and have strong potential for a measurement. 
The existing LEP data are publicly available through the CERN Open Data portal and could, in principle, also be used to search for NAQC.
However, the available statistics are considerably smaller, and a dedicated extraction of the NAQC signal from these data is beyond the scope of the present work.

\begin{table}[t]
\centering
\begin{tabular}{lccc}
\hline\hline
Experiment & COM [GeV] & $\mathcal L$ & $N_{NAQC}$ \\
\hline
Belle & 10.58 & 1~ab$^{-1}$ & $2\cdot10^6$ \\
Belle~II & 10.58 & 50~ab$^{-1}$ & $8\cdot10^7$ \\
FCC-$ee$ & 91.19 & 125~ab$^{-1}$ & $8\cdot10^8$ \\
LEP & $\sim 91.19$ & 116~pb$^{-1}$ & $7\cdot10^2$ \\
\hline\hline
\end{tabular}
\caption{Approximate number of $\tau^+\tau^- \to \pi^+\bar{\nu}_{\tau}\pi^-\nu_{\tau}$ events available in the phase-space regions exhibiting NAQC for selected experiments.}
\label{tab:nevents}
\end{table}

\subsection{Experimental steering game}\label{subsec:SteeringGame}

The above proposal based on the measurement of an NAQC witness provides an experimental certification of the ability of the underlying quantum state to display NAQC. However, following Ref.~\cite{Afik:2022dgh}, we also design an experimental scheme to implement the actual steering game leading to the definition of NAQC, see Eqs.~(\ref{eq:ConditionalProbability})-(\ref{eq:NAQCDef}) and ensuing discussion. For that purpose, we restrict once more to $\tau^+\tau^- \to \pi^+\bar{\nu}_{\tau}\pi^-\nu_{\tau}$ decays as the maximal spin analyzing power of pions is now essential. The associated normalized cross section, Eq.~(\ref{eq:TauCrossSectionNormalized}), leads to the following probability distribution for the $\pi^\mp$ flight directions $\mathbf{\hat{q}}_{\pm}$ in their respective parent rest frames:
\begin{equation}
p(\mathbf{\hat{q}}_{+},\mathbf{\hat{q}}_{-})=
\frac{1+\mathbf{B}^{+}\cdot\mathbf{\hat{q}}_{+}-\mathbf{B}^{-}\cdot\mathbf{\hat{q}}_{-}-\mathbf{\hat{q}}_{+}\cdot \mathbf{C} \cdot\mathbf{\hat{q}}_{-}}{(4\pi)^2}.
\end{equation}
The marginal distributions for the flight direction of each pion are given by
\begin{equation}
p(\mathbf{\hat{q}}_{\pm})=\int\mathrm{d}\Omega_{\mp}~p(\mathbf{\hat{q}}_{+},\mathbf{\hat{q}}_{-})=\frac{1\pm \mathbf{B}^{\pm}\cdot\mathbf{\hat{q}}_{\pm}}{4\pi},
\end{equation}
from where we can measure the probability in Eq.~(\ref{eq:ConditionalProbability}) as
\begin{equation}
p_{\mathbf{\hat{n}}}=\frac{p(\mathbf{\hat{q}}_{-}=-\mathbf{\hat{n}})}{p(\mathbf{\hat{q}}_{-}=\mathbf{\hat{n}})+p(\mathbf{\hat{q}}_{-}=-\mathbf{\hat{n}})}.
\end{equation}
Furthermore, we can fit the conditional Bloch vector, Eq.~(\ref{eq:ConditionalBloch}), from the conditional probability distribution
\begin{equation}
\label{eq:ConditionalTomography}
p(\mathbf{\hat{q}}_{+}|\mathbf{\hat{q}}_{-}=- \mathbf{\hat{n}})=\frac{p(\mathbf{\hat{q}}_{+},\mathbf{\hat{q}}_{-}=- \mathbf{\hat{n}})}{p(\mathbf{\hat{q}}_{-}=-\mathbf{\hat{n}})}=\frac{1+ \mathbf{B}_{\mathbf{\hat{n}}}^{+}\cdot\mathbf{\hat{q}}_{+}}{4\pi}.
\end{equation}
Thus, we can reconstruct the six conditional quantum states $\rho^a_j$ and measure the associated probabilities $p^a_j$ in Eq.~(\ref{eq:NAQCDef}), with $a=\pm1,\,j=1,2,3$, by taking $\mathbf{\hat{n}}=a \mathbf{\hat{n}}_j$ in the above expressions, $\mathbf{\hat{n}}_j$ being each of the vectors of the chosen orthonormal basis. 
Once each state $\rho^a_j$ has been reconstructed, the coherence in the remaining directions $i\neq j$ is measured [e.g., by direct evaluation of $\mathfrak{C}_i^{l_1}(\rho^a_j)$]. 
By averaging over all possible directions, the NAQC measure~(\ref{eq:NAQCDef}) can be evaluated. 

The proposed scheme thus provides a direct high-energy counterpart of the photonic experiment in Ref.~\cite{Ding2019}.
Here, the crucial step is the possibility of reconstructing directly the conditional quantum states from Eq.~(\ref{eq:ConditionalTomography}), as originally discussed in Ref.~\cite{Afik:2022dgh}. 
Specifically, we propose to use the diagonal basis for the steering game in low-energy colliders, such as Belle~II, while the helicity basis can be used in high-energy colliders, such as FCC-$ee$, since there the diagonal basis is equivalent to the helicity basis to a very good approximation. 
An experimental protocol for spin measurements in the diagonal basis is developed in Appendix~\ref{app:Experimental}. 
The above steering game can be easily inverted to measure the conditional states on the $\tau^+$ side.
Nevertheless, a full dedicated analysis of the implementation of the steering game, including an estimation of the associated uncertainties, is beyond the scope of this work and left for the future.

\section{Conclusions and outlook}\label{sec:Conclu}

In this work, we quantify the presence of NAQC in $e^+e^- \to \tau^+\tau^-$ using two measures of coherence: the $l_1$-norm of coherence and the relative entropy of coherence. 
We find that the $l_1$-norm captures NAQC over a larger region of phase space and therefore provides the most promising observable for an experimental measurement. 
We also develop simple analytical models that enable a rapid and accurate calculation of NAQC across different experimental configurations, providing a useful tool for assessing the prospects of its observation at present and future collider experiments.
In particular, we formulate an NAQC witness which allows to certify the presence of NAQC from the measurement of a single magnitude. 
We then focus on two benchmark COM energies, $\sqrt{s}=10.58$~GeV and $91.19$~GeV, corresponding to the Belle~II and FCC-$ee$ programs, respectively. 
We find both setups quite promising for a potential NAQC observation, as a fraction of 0.18 (Belle~II) and 0.31 (FCC-$ee$) of the total number of events can display NAQC.
We further quantify the experimental precision required to establish the presence of NAQC, providing a concrete target for future measurements.
Our results indicate that FCC-$ee$ offers the greatest sensitivity to NAQC, while Belle~II also provides a promising avenue for its observation.
Finally, we propose an experimental steering game in a high-energy collider that allows to measure NAQC according to its very definition.

Our work further establishes NAQC as the strongest form of quantum correlations studied to date in collider physics. 
In general, NAQC is a sufficient condition for entanglement and steerability~\cite{Mondal2017} and, for states with diagonal correlation matrix, also for Bell nonlocality~\cite{Hu2018}. 
Moreover, there is strong numerical evidence~\cite{Hu2018} suggesting the following hierarchy extension for two qubits:
\begin{equation*}
\begin{aligned}
\textrm{NAQC} \subset \textrm{Bell Nonlocality} \subset \textrm{Steering} \subset 
\\ \textrm{Entanglement} \subset \textrm{Discord}.
\end{aligned}
\end{equation*}
Our results provide another layer of confirmation for this conjecture using high-energy qubits, going also beyond the hierarchy experimentally characterized in top quarks~\cite{Afik:2026pxv}. The observation of NAQC would thus constitute another milestone in the field of Quantum Information in High-Energy Physics, representing the strongest quantum correlation yet observed in a high-energy collider experiment. It would also provide a high-energy complementary measurement to the NAQC observation of Ref.~\cite{Ding2019}, which still remains the only NAQC experimental observation in the literature to the best of our knowledge.

In addition, our proposal for a steering game would represent the first implementation of a steering game in a high-energy collider, paving the way for the realization of more general steering games. 
Within Quantum Information, steering games provide a fundamental operational framework to investigate and characterize quantum resources~\cite{Cavalcanti2013,Banik2013,Sun2014,Kocsis2015,Gheorghiu2017}.
Apart from further bridging Quantum Information and High-Energy Physics, the asymmetric nature of steering games makes them useful quantum tools to probe $CP$-violating physics beyond the Standard Model, a search program originally proposed in Ref.~\cite{Afik:2022dgh} and strongly advocated recently in Ref.~\cite{Alvarez2026}.

Several directions remain open for future study. 
The methodology developed in this work is readily applicable to other present and future electron-positron colliders. 
It can be also adapted to other electroweak annihilation processes involving $\gamma/Z$ mediation.
More generally, it would be interesting to investigate NAQC in other systems, beyond $e^+e^- \to\tau^+\tau^-$ and $pp \to t\bar t$. 
Future work should perform a dedicated experimental analysis of our proposal for the implementation of a steering game as well as investigate alternative steering games that can be implemented in collider experiments. 
Finally, another intriguing research line would be the exploration of how physics beyond the Standard Model could modify the structure of NAQC and other steering games. 
All these studies establish NAQC as a new tool for probing both fundamental quantum phenomena and particle-physics dynamics at high energies.

\acknowledgments{We thank Luca Marzola for advising on Monte Carlo and theoretical calculations. YA is supported by the National Science Foundation under Grant No.\ PHY-2310094. JRMdN acknowledges funding from Spain's MICIU/AEI through Ram\'on y Cajal program (Grant No. RYC2024-050437-I) and through Proyectos de Generaci\'on de Conocimiento (Grant No. PID2025-169103NA-I00). SSR acknowledges funding recieved from University of British Columbia's Four-Year Fellowship.}

\bibliography{main.bib}

\appendix

\section{Derivation of NAQC witness}\label{app:witness}

Our starting point is Eq.~(\ref{eq:NAQCExplicit}). We first notice that
\begin{equation}
\frac{1}{2}\sum_{a=\pm 1}|\mathbf{\hat{n}}_i\times\mathbf{B}^+ +a\mathbf{\hat{n}}_i\times(\mathbf{C}\cdot\mathbf{\hat{n}}_j)|\geq |\mathbf{\hat{n}}_i\times(\mathbf{C}\cdot\mathbf{\hat{n}}_j)|,
\end{equation}
where we have used the triangle inequality
\begin{equation}
|\mathbf{x}|=\frac{{}|(\mathbf{x}-\mathbf{y})+(\mathbf{x}+\mathbf{y})|}{2}\leq \frac{|\mathbf{x}-\mathbf{y}|+|\mathbf{x}+\mathbf{y}|}{2}.
\end{equation}
As a result,
\begin{align}
\frac{1}{2} \sum^3_{j=1}\sum^3_{\substack{i=1 \\ i\neq j}}|\mathbf{\hat{n}}_i\times(\mathbf{C}\cdot\mathbf{\hat{n}}_j)|\leq \mathfrak{C}_{\rm{NA}}^{l_1}(\rho).
\end{align}
In addition,
\begin{equation}
\mathbf{C}\cdot\mathbf{\hat{n}}_j=\sum^3_{k=1}C_{kj}\mathbf{\hat{n}}_k,
\end{equation}
which, for $i\neq j$, yields
\begin{equation}
|\mathbf{\hat{n}}_i\times(\mathbf{C}\cdot\mathbf{\hat{n}}_j)|=\sqrt{\sum^3_{\substack{k=1 \\ k\neq i}}C^2_{kj}}\geq |C_{jj}|.
\end{equation}
Hence, we arrive at
\begin{equation}
\label{eq:WitnessInequality}
W_{\rm{NA}}(\rho)\equiv\sum^3_{i=1}|C_{ii}|\leq\frac{1}{2} \sum^3_{j=1}\sum^3_{\substack{i=1 \\ i\neq j}}|\mathbf{\hat{n}}_i\times(\mathbf{C}\cdot\mathbf{\hat{n}}_j)|\leq \mathfrak{C}_{\rm{NA}}^{l_1}(\rho).
\end{equation}
In the case where $\mathbf{C}$ is symmetric, the NAQC witness is maximized in the diagonal basis since 
\begin{equation}
\label{eq:WitnessDiagonal}
\sum^3_{i=1}|C_{ii}|\leq \sum^3_{i=1}|C_{i}|.
\end{equation}
This can be proven by invoking Von Neumann trace inequality~\cite{Mirsky1975},
\begin{equation}
\left|\textrm{Tr}[AB]\right|\leq \sum^n_{i=1}\alpha_i\beta_i,
\end{equation}
where $A,B$ are complex $n\times n$ matrices with singular values $0\leq \alpha_n\leq \ldots \leq \alpha_2\leq \alpha_1$ and $0\leq \beta_n\leq \ldots \leq \beta_2\leq \beta_1$, respectively. Then, one simply takes $A$ with matrix elements $A_{ij}=\textrm{sign}(C_{ij})\delta_{ij}$ and $B=\mathbf{C}$. In that case, $\alpha_i=1$, and
\begin{equation}
\left|\textrm{Tr}[AB]\right|=\sum^3_{i=1}|C_{ii}|\leq \sum^3_{i=1}\alpha_i\beta_i=\sum^3_{i=1}\beta_i=\sum^3_{i=1}|C_i|.
\end{equation}

\section{Analytical expressions for the spin density matrix}\label{app:formulas}

We provide here the leading-order expressions entering the computation of the spin density matrix of $\tau^+\tau^-$ pairs produced via $e^+e^-$ annihilation; see Refs.~\cite{Fabbrichesi:2022ovb,Maltoni2024,Han:2025ewp,Zhang:2025mmm} for more details. First, the spin-summed rate function $c_0$ determining the differential cross section [see Eq.~(\ref{eq:dsigma})] is 
\begin{equation}
\label{eq:c0}
c_0=F_0\left(2-\beta^2\sin^2\Theta\right)+2F_1\cos\Theta+F_2\left(1+\cos^2\Theta\right),
\end{equation}
with the form factors $F_{0,1,2}$ given by
\begin{widetext}
\begin{align}
\label{eq:FormFactors}
\nonumber F_0&=48\Big[Q_\tau^2Q_e^2+2Q_\tau Q_eg_V^\tau g_V^er_Z+(g_V^\tau)^2 \big[(g_V^e)^2+(g_A^e)^2\big]d_Z\Big],
\\
F_1&=48g_A^\tau g_A^e\beta\left[2Q_\tau Q_er_Z+4g_V^\tau g_V^ed_Z\right],\\
\nonumber F_2&=48(g_A^\tau)^2\beta^2\big[(g_V^e)^2+(g_A^e)^2\big]d_Z,
\end{align}
where $Q_f$, $g_A^f=T_3^f/2$, and $g_V^f=g_A^f-Q_f\sin^2\theta_W$ are the electric charge and $Z$ couplings of fermion $f$, with $T_3^f$ its weak isospin. Specifically, the fermion electric charges and weak isospins are $Q_\tau=Q_e=-1$ and $T^\tau_3=T^e_\tau=-1/2$, so $g_A^\tau=g_A^e=g_A=-1/4$ and $g_V^\tau=g_V^e=g_V=-1/4+\sin^2\theta_W$. The above expressions then simplify to
\begin{align}
\label{eq:FormFactorsSimple}
\nonumber F_0&=48\Big[1+2g_V^2r_Z+g_V^2 (g_V^2+g_A^2)d_Z\Big],
\\
F_1&=48g_A^2\beta\left[2r_Z+4g_V^2 d_Z\right],\\
\nonumber F_2&=48g_A^2\beta^2(g_V^2+g_A^2)d_Z.
\end{align}

The pure photon contribution is encoded in the first term between brackets for the expression of $F_0$. The $\gamma$--$Z$ interference and pure $Z$-boson production contribute, respectively, through the terms
\begin{equation}
r_Z\equiv \frac{{\rm Re}(G_Z)}{|G_Z|^2},\qquad d_Z\equiv\frac{1}{|G_Z|^2},
\end{equation}
which arise from the dimensionless propagator factor
\begin{equation}
\label{eq:DZ}
G_Z=\sin^2\theta_W\cos^2\theta_W\frac{(s-m_Z^2)+im_Z\Gamma_Z}{s},
\end{equation}
$\theta_W$ being the weak mixing angle, and $m_Z$, $\Gamma_Z$ the $Z$-boson mass and width, respectively. 

The elements of the spin-correlation matrix in the helicity basis read, in term of these form factors, as
\begin{align}
\label{eq:SpinCorrelations}
\nonumber C_{kk}&=\frac{F_0\left[\beta^2+(2-\beta^2)\cos^2\Theta\right]+2F_1\cos\Theta+F_2(1+\cos^2\Theta)}{c_0},\\
C_{rr}&=\frac{\left[F_0(2-\beta^2)-F_2\right]\sin^2\Theta}{c_0},\\
\nonumber C_{nn}&=\frac{\left(F_2-F_0\beta^2\right)\sin^2\Theta}{c_0},\\
\nonumber C_{kr}&=-\,\sqrt{1-\beta^2}\;\frac{\left(2F_0\cos\Theta+F_1\right)\sin\Theta}{c_0}.
\end{align}
On the other hand, the components of the polarization vectors read
\begin{align}
\label{eq:Polarization}
\nonumber B_k^{\pm}&=B_k=-\frac{F_3(1+\cos^2\Theta)+2F_4\cos\Theta}{c_0},\\
B_r^{\pm}&=B_r=\sqrt{1-\beta^2}\frac{(F_3\cos\Theta+2F_5)\sin\Theta}{c_0},
\end{align}
where we have defined the following form factors for compactness:
\begin{align}
\label{eq:FormFactorsPol}
\nonumber F_3&\equiv96\beta\Big[Q_\tau Q_eg_A^\tau g_V^er_Z+g^\tau_Ag_V^\tau \big[(g_V^e)^2+(g_A^e)^2\big]d_Z\Big],
\\
F_4&\equiv96\Big[Q_\tau Q_eg_A^e g_V^\tau r_Z+g^e_Ag_V^e \big[(g_V^\tau)^2+\beta^2(g_A^\tau)^2\big]d_Z\Big],\\
\nonumber F_5&\equiv 96\Big[Q_\tau Q_eg_A^e g_V^\tau r_Z+g^e_Ag_V^e (g_V^\tau)^2d_Z\Big].
\end{align}

\end{widetext}

In the same spirit of Eq.~(\ref{eq:FormFactorsSimple}), these expressions can be simplified to
\begin{align}
\label{eq:FormFactorsPolSimple}
\nonumber F_3&=96g_Ag_V\beta\Big[r_Z+\big(g_V^2+g_A^2\big)d_Z\Big],
\\
F_4&=96g_Ag_V\Big[r_Z+ \big(g_V^2+\beta^2g_A^2\big)d_Z\Big],\\
\nonumber F_5&=96g_Ag_V\Big[r_Z+g_V^2d_Z\Big].
\end{align}
Notice that the emergence of non-vanishing spin polarizations is an exclusive feature of the weak interaction, as they only depend on $r_Z,d_Z$.

The numerical values used in this work are $m_\tau=1.777$~GeV, $m_Z=91.1876$~GeV, $\Gamma_Z=2.4952$~GeV and $\sin^2\theta_W=0.2312$. 

\section{Approximate models}\label{app:App}

We examine here different limits of the analytical expressions provided in the previous section.

\subsubsection{QED production}

For $\sqrt{s}\ll m_Z$, the photon channel dominates over the $Z$ channel, and then $F_0\simeq 48\gg |F_{n}|$, $n\neq 0$. Specifically, at the Belle~II COM energy, $\sqrt{s}=10.58$~GeV ($\beta\approx 0.9419$), we have $r_Z\approx -0.077$ and $d_Z\approx 0.0059$. This yields $F_0\approx 47.997$ and
\begin{equation}
\label{eq:FormFactorsBelle}
F_1\approx -0.4,~ F_2\approx 10^{-3},~F_3\approx F_4\approx F_5\approx -0.03.
\end{equation}
Thus, we can set $F_0=48$ and $F_n=0$ for $n\neq 0$ in Eqs.~(\ref{eq:c0}),~(\ref{eq:SpinCorrelations}),~(\ref{eq:Polarization}). This results in the $T$-state of Eq.~(\ref{eq:SimpleBelle}) of the main text.

\subsubsection{Ultrarelativistic limit}

The $Z$-boson contribution becomes relevant for energies $\sqrt{s}\gtrsim m_Z$. Nevertheless, since $m_Z\gg m_\tau$, we can work here in the ultrarelativistic limit $\beta \simeq 1$, where
\begin{align}
\label{eq:Ultrarelativistic}
c_0&\simeq (F_0+F_2)\left(1+\cos^2\Theta+2A\cos\Theta\right),\\
\nonumber C_{kk}&\simeq 1,\\
\nonumber C_{nn}&\simeq -C_{rr}\simeq N\frac{\sin^2\Theta}{1+\cos^2\Theta+2A\cos\Theta}\equiv C,\\
\nonumber B_{k}&\simeq-B\frac{(1+\cos\Theta)^2}{1+\cos^2\Theta+2A\cos\Theta},
\\
\nonumber C_{kr}&\simeq B_{r}\simeq 0,
\end{align} 
with
\begin{equation}
N\equiv \frac{F_2-F_0}{F_2+F_0},~A\equiv\frac{F_1}{F_2+F_0},~B\equiv\frac{F_3}{F_2+F_0}\simeq\frac{F_4}{F_2+F_0}.
\end{equation}
As a result, in the ultrarelativistic limit, the helicity basis is equivalent to the diagonal basis, and the state is polarized solely along the $\hat{k}$ axis. In particular, the coefficient $N$ controls the magnitude of the transverse spin correlations $C$, while the coefficient $B$ controls the magnitude of the longitudinal spin polarization. The coefficient $A$ is another genuine electroweak feature, characterizing the forward-backward asymmetry in $\tau^+\tau^-$ production.

For the state given by Eq.~(\ref{eq:Ultrarelativistic}), using Eq.~(\ref{eq:NAQCExplicit}), we find that
\begin{equation}
\mathfrak{C}^{l_1}_{\rm{NA}}=1+|C|+\sqrt{C^2+B_k^2}\geq 1+2|C|=W_{\rm{NA}}.
\end{equation}
Thus, NAQC is mainly controlled by the behavior of $C$. It is easy to see that, for given COM energy, $C$ is peaked at
\begin{equation}
\label{eq:MaxCorrelations}
\cos\Theta=-\frac{1}{A}+\textrm{sign}(A)\cdot\sqrt{\frac{1}{A^2}-1},
\end{equation}
where we stress that $A$ only depends on $\sqrt{s}$. The resulting curve in phase space accurately predicts the center of the NAQC bands (white line in Figure~\ref{fig:naqc_maps}).

On the other hand, $C$ vanishes when $N$ vanishes, i.e., when $F_0=F_2$. This occurs for
\begin{equation}
G_Z\simeq-g^2_V\pm g^2_A\simeq \pm g^2_A=\pm \frac{1}{16},
\end{equation}
where we have used that, outside the $Z$-peak, $G_Z\simeq \textrm{Re}(G_Z)$, and that $g^2_V\approx 3\cdot10^{-4}\ll g^2_A=1/16$. Consequently, we find that $C=0$ at the COM energies 
\begin{equation}
\label{eq:MinimalCorrelations}
\sqrt{s}_{\pm}\simeq \frac{m_Z}{\sqrt{1\mp\dfrac{1}{16\sin^2\theta_W\cos^2\theta_W}}},
\end{equation}
which yields $\sqrt{s}_{-}\approx 78~\textrm{GeV}$ and $\sqrt{s}_{+}\approx 113~\textrm{GeV}$. These two values correspond to the two gaps separating the three NAQC bands observed in Figure~\ref{fig:naqc_maps} (horizontal solid lines).

\subsubsection{$Z$-peak}

Further approximations on top of the ultrarelativistic limit can be made at the $Z$-peak, $\sqrt{s}=m_Z$, where $\beta\approx 0.9992 \simeq 1$. This is because there
\begin{equation}
r_Z=0,\quad d_Z=\frac{1}{\sin^4\theta_W\cos^4\theta_W}\frac{m^2_Z}{\Gamma^2_Z}\approx 4.23\cdot 10^4 \gg 1.
\end{equation}
As a result, one can only retain the $d_Z$ contribution in Eqs.~(\ref{eq:FormFactorsSimple}),~(\ref{eq:FormFactorsPolSimple}), resulting in
\begin{align}
\nonumber N&\simeq\frac{g^2_A-g^2_V}{g^2_A+g^2_V}\approx 0.989,\\
A&\simeq B^2\approx 0.022,\\
\nonumber B&\simeq\frac{2g_Ag_V}{g^2_A+g^2_V}\approx 0.150.
\end{align}
Since $A\ll 1$ and we focus on the signal close to the NAQC peak around $\cos\Theta=0$, we can safely take $A=0$ in Eq.~(\ref{eq:Ultrarelativistic}), arriving at
\begin{align}
\label{eq:UltrarelativisticFCC}
c_0&\simeq (F_0+F_2)(1+\cos^2\Theta),\\
\nonumber C_{kk}&=1,\\
\nonumber C&\simeq N\frac{\sin^2\Theta}{1+\cos^2\Theta},\\
\nonumber B_{k}&\simeq-B\frac{(1+\cos\Theta)^2}{1+\cos^2\Theta}.
\end{align} 
We note that, at the $Z$-peak, $C_{kr}$ and $B_r$ are suppressed by both the high energy $\sqrt{1-\beta^2}=2m_\tau/\sqrt{s}\simeq0.039$ and the form factors ratio $F_n/F_2\ll 1$ for $n\neq 2$. Specifically, 
\begin{equation}
\label{eq:FormFactorsFCC}
F_0\approx 93,~F_1\approx 179,~F_2\approx 7959,~F_3\approx F_4\approx 1197,~F_5\simeq 7.
\end{equation}
Remarkably, when performing the symmetric angular average of Eq.~(\ref{eq:PolarQuantumState}), even when working with the full ultrarelativistic state (\ref{eq:Ultrarelativistic}), one arrives at the same result for the averaged spin coefficients as the contribution from odd terms in $\cos\Theta$ vanishes:
\begin{align}
\label{eq:UltrarelativisticFCCAngular}
c_0(\sqrt{s},x_c)&=(F_0+F_2)\left(1+\dfrac{x_c^2}{3}\right),\\
\nonumber C_{kk}(\sqrt{s},x_c)&=1,\\
\nonumber C_{rr}(\sqrt{s},x_c)&=-C_{nn}(\sqrt{s},x_c)=N\frac{1-\dfrac{x_c^2}{3}}{1+\dfrac{x_c^2}{3}},\\
\nonumber B_{k}(\sqrt{s},x_c)&=-B.
\end{align} 
Finally, the simple model of Eq.~(\ref{eq:SimpleFCC}) amounts to take $N=1$ and $B=0$ in Eq.~(\ref{eq:UltrarelativisticFCC}).

\section{Experimental protocol for spin measurements in the diagonal basis}\label{app:Experimental}

As explained in the main text, see Eq.~(\ref{eq:eigenvalues}) and ensuing discussion, the diagonal basis is given by the eigenvectors $\{\hat{u}_{i}\}^3_{i=1}=\{\hat{u}_{+},\hat{u}_{-},\hat{n}\}$ of the spin-correlation matrix, satisfying
\begin{equation}
\mathbf{C}\cdot\hat{u}_{i}=C_i\hat{u}_{i}.
\end{equation}
Specifically, the correlation matrix can be diagonalized by a certain rotation in the $\{\hat{k},\hat{r}\}$ plane, 
\begin{align}
\label{eq:DiagonalBasisRotation}
\hat{u}_{+}&=\hat{k}\cos\alpha+\hat{r}\sin\alpha,\\
\nonumber \hat{u}_{-}&=-\hat{k}\sin\alpha+\hat{r}\cos\alpha,
\end{align}
so
\begin{equation}
\mathbf{R}=\left[\begin{array}{ccc}
\cos\alpha &-\sin\alpha & 0 \\
\sin\alpha & \cos\alpha & 0\\
0 & 0 & 1
\end{array}\right],~\mathbf{R}\trans\mathbf{C}\mathbf{R}=\left[\begin{array}{ccc}
C_{+} & 0 & 0 \\
0 & C_{-} & 0\\
0 & 0 & C_{nn}
\end{array}\right].
\end{equation}
The angle $\alpha$ depends on $(\sqrt{s},\cos\Theta)$ through the spin-correlation coefficients in Eq.~(\ref{eq:SpinCorrelations}) as
\begin{equation}
\label{eq:alphadiagonal}
\alpha=\frac{1}{2}\arcsin\frac{C_{kr}}{\sqrt{\left(\dfrac{C_{kk}-C_{rr}}{2}\right)^{2}+C_{kr}^{2}}}.
\end{equation}

Experimentally, in order to measure the spin coefficients in the diagonal basis from Eq.~(\ref{eq:TauCrossSectionNormalized}), instead of determining event by event the cosines of the product directions $\mathbf{\hat{q}}_{\pm}$ with respect to the helicity basis, one determines their cosines with respect to the diagonal basis,
\begin{equation}
\label{eq:directorcosines}
\cos\theta^\pm_i=\mathbf{\hat{q}}_{\pm}\cdot \hat{u}_{i}.
\end{equation}
The vectors in the above expression are evaluated in the respective parent rest frames after boosting from the COM frame, where we recall that $\mathbf{\hat{q}}_{\pm}$ corresponds to the selected decay product of the $\tau^{\mp}$ lepton. In particular, notice that the helicity basis is not deformed when boosting from the COM frame to each parent $\tau^{\mp}$ frame since the boost is collinear with $\hat{k}$. Hence, we can intrinsically determine $\hat{u}_{i}=\hat{u}_{\pm}$ in each parent frame using Eq.~(\ref{eq:DiagonalBasisRotation}), with $\alpha$ evaluated according to the kinematics of the underlying event.

In general, any vector $\mathbf{A}$ given in the helicity basis (e.g., the spin polarization) is transformed into a vector $\mathbf{A}'$ in the diagonal basis by applying the transpose of the rotation matrix:
\begin{equation}
\mathbf{A}'=\mathbf{R}\trans\cdot \mathbf{A}.
\end{equation}

In summary, for measuring the spin coefficients in the diagonal basis, one just needs to evaluate the angle $\alpha$ analytically from Eq.~(\ref{eq:alphadiagonal}) and perform the rotation of Eq.~(\ref{eq:DiagonalBasisRotation}) in the respective parent rest frames. Once done, the rest is the usual procedure to obtain the spin correlations and polarizations using the angular distributions associated to the cosines from Eq.~(\ref{eq:directorcosines}).

\end{document}